\documentclass{aa}

\usepackage{graphicx}
\usepackage{multirow}
\usepackage{txfonts}
\usepackage{natbib}
\usepackage{hyperref}
\usepackage{float}
\hypersetup{
    colorlinks=true,
    linkcolor=magenta,
    citecolor=blue,
    urlcolor=magenta,
    }

\newcommand{\halpha}{\text{H}\alpha}
\newcommand{\hbeta}{\text{H}\beta}    
\newcommand{\nii}{[\text{NII}]\lambda6584}
\newcommand{\oiii}{[\text{OIII}]\lambda5007}
\newcommand{\sii}{[\text{SII}]\lambda\lambda6717,6731}

\bibpunct{(}{)}{;}{a}{}{,}

\begin{document}

\title{Through the fog: a complementary optical galaxy classification scheme for `intermediate' redshifts}
\author{Duarte Muñoz Santos\inst{1,2} \and Cirino Pappalardo \inst{2,3} \and Henrique Miranda \inst{2,3} \and José Afonso \inst{2,3} \and Israel Matute \inst{2,3} \and Rodrigo Carvajal \inst{2,3} \and Catarina Lobo \inst{4,5} \and Patricio Lagos\inst{4,6} \and Polychronis Papaderos\inst{2,3,4} \and Ana Paulino-Afonso\inst{4} \and Abhishek Chougule\inst{4,5} \and Davi Barbosa \inst{2,3} \and Bruno Lourenço \inst{3}}

\institute{Aix Marseille Univ, CNRS, CNES, LAM, Marseille, France \\ E-mail: \href{mailto:duarte.santos@lam.fr}{\texttt{duarte.santos@lam.fr}} \and Instituto de Astrofísica e Ciências do Espaço, Universidade de Lisboa - OAL, Tapada da Ajuda, PT1349-018 Lisboa, Portugal \and Departamento de Física, Faculdade de Ciências da Universidade de Lisboa, Edifício C8, Campo Grande, PT1749-016 Lisboa, Portugal \and Instituto de Astrofísica e Ciências do Espaço, Universidade do Porto - CAUP, Rua das Estrelas, PT4150-762 Porto, Portugal \and Departamento de Física e Astronomia, Faculdade de Ciências, Universidade do Porto, Rua do Campo Alegre 687, PT4169-007 Porto, Portugal \and Institute of Astrophysics, Facultad de Ciencias Exactas, Universidad Andr\'es Bello, Sede Concepci\'on, Talcahuano, Chile}

\date{Received ?? \ Accepted ??}

\abstract{Understanding galaxy evolution strongly depends on our interpretation of their spectra. In the optical, the BPT diagrams have been the main way to distinguish if the principal excitation mechanism of a galaxy is dominated by star-formation (SF) or an Active Galactic Nucleus (AGN). Although different classification methods exist, these are based on either hard to detect or high-energy emission lines. To date, the Balmer lines remain the most consistent way to classify galaxies, but at `intermediate' redshifts ($1.5 < z < 2.5$), galaxies are hard to parse in the BPT diagrams (and siblings) because the crucial $\halpha$ emission line is out of the range of ground-based optical spectrographs.}{In this work, we re-explore a diagram, which we call the OB-I diagram, that compares the equivalent width of $\hbeta$ and the emission line flux ratio of $\oiii/\hbeta$ and breathe new life into it, as it has the potential to be used for the classification of galaxies at these `intermediate' redshifts and `illuminate the fog' that permeates galaxy classification in the optical restframe.}{We use data from SDSS, LEGA-C, VANDELS, JADES, 3D-HST and MOSDEF to explore galaxy classification in the OB-I diagram across a wide range of redshifts ($0 < z < 2.7$).}{Our results show that, at $z < 0.4$, the OB-I diagram clearly separates galaxies between two distinct types: one dominated by an AGN and a second made up of a mixed population of SF galaxies and AGN activity. Comparison with the BPT diagrams and theoretical models shows that this mixed population can be partially separated from a pure SF population, with a simple semi-empirical fit. At higher redshifts, we find that half of AGNs identified by other classification schemes are correctly recovered by the OB-I diagram, potentially making this diagram resistant to the `cosmic shift' that plagues most optical classification schemes, but more research is needed to understand this phenomenon.}{We find the OB-I diagram, which only requires two emission lines to be implemented, to be a useful tool at separating galaxies that possess a dominating AGN component in their emission from others. This applies not only to the Local Universe, but also seemingly at redshifts near the Cosmic Noon ($z \sim 2$), without any need for significant adjustments in our empirical fit.}


\maketitle

\section{Introduction}\label{section:intro}

Galaxies are complex and dynamic systems, and analysing their spectra is one of the best tools we have to better comprehend them. To understand the nature of galaxies, however, there is a need to distinguish if their principal excitation mechanisms are governed by photoionisation from hot, massive stars - directly related with ongoing or recent star-forming (SF) activity \citep{osterbrock} - or from an active galactic nucleus (AGN).

Five decades ago, \cite{searle72} noticed that line ratios were efficient at separating older from younger spiral galaxies. In the 1980s, \cite{bpt} took advantage of this to create several diagrams that compared the ratio of specific emission lines and were able to separate galaxies according to their main excitation mechanism. Further investigation of these diagrams by \cite{veilleux} showed that there were three diagrams that performed this efficiently: $[\text{NII}]\lambda6584/\text{H}\alpha$ vs. $[\text{OIII}]\lambda5007/\text{H}\beta$; $[\text{SII}]\lambda\lambda6717,6731/\text{H}\alpha$ vs. $[\text{OIII}]\lambda5007/\text{H}\beta$ and $[\text{OI}]\lambda6300/\text{H}\alpha$ vs. $[\text{OIII}]\lambda5007/\text{H}\beta$ diagrams, henceforth to be known as the NII, SII and OI diagrams, respectively. To simplify, galaxies were separated into two clear and distinct types: SF and AGNs.

Nearly twenty years later, thanks to projects like the Sloan Digital Sky Survey (SDSS, \citealt{sdss_technical1,sdss_technical2,sdss_technical3}) the interpretation of these diagrams was refined. In \cite{kewley_2001}, through a combination of stellar population synthesis and photoionisation models, the authors were able to determine the first theoretical `maximum starburst' line in the NII, SII and OI diagrams, which determined the location of galaxies in the parameter space whose emission is dominated by star-forming activity. Furthermore, \cite{kauffmann_2003} adjusted a semi-empirical fit to the NII diagram, which allowed for the definition of another class: Composite galaxies, or galaxies that have a mix of both star-formation and AGN activity. This separation between SF and Composite galaxies is due to the fact that the $[\text{NII}]\lambda6584/\text{H}\alpha$ ratio saturates at high metallicities, considering only the nebular contribution \citep{kewley_2002,denicolo_2002}. For galaxies to have higher contributions of the $[\text{NII}]\lambda6584$ emission line, there must be a mechanism inside the galaxy that boosts the value of its flux - in this case, an AGN \citep{kewley}.

The BPT diagrams have been proven time and time again to be a great tool that astrophysicists can use to separate galaxy types, and a countless number of papers use it to classify galaxies and explore the consequences of these classes (e.g. \citealt{patricio,polimera22,harish23,teimoorinia24}, to cite just a few). However, it is not the only classification scheme available using optical lines. Another relevant example is the WHAN diagram \citep{cidfernandes}, named as such because it uses the equivalent width (EW) of $\halpha$ as well as the flux ratio of $\halpha$ and $\nii$. This classification scheme is mostly used to distinguish active galaxies (galaxies that are actively forming stars) from passive ones (galaxies that show no star-formation in the present). Instead of Balmer lines, one can base the classification on oxygen or nitrogen lines \citep{Rola97,paalvast,perrotta}. There are also the mass-excitation diagrams \citep{juneau11}, which combine the emission lines of galaxies with their physical properties in order to distinguish the effects and degeneracies present in these ratios.

All of these diagrams are useful but they are mostly applied in the Local Universe ($z < 0.1$). At higher redshifts, the BPT diagram becomes difficult to interpret, but in the advent of advanced space telescopes such as the James Webb Space Telescope (JWST, \citealt{jwst}), diagrams that use higher-order emission lines (e.g. forbidden line ratio, FLR, \citealt{flr}; comparison between oxygen, hydrogen and neon lines, OHNO, \citealt{ohno}) and high-energy or auroral lines (e.g. HeII/$\hbeta$, \citealt{heii}; $[\text{OIII}]\lambda4363/\text{H}\gamma$, \citealt{mazzolari24}) are now able to be plotted for a wide selection of galaxies, providing the scientific community with new tools to separate galaxy types and disentangle their properties.

However, there is a lot of data that does not reside in the realm of detection of space-based instruments like JWST. Efforts like the aforementioned SDSS and the Dark Energy Spectroscopic Instrument (DESI, \citealt{desi_edr}) surveys reside in the optical and, in this regime, it is not always possible to apply the previously referred diagnostics either due to: 1) emission lines such as $\halpha$ moving out of the range of the spectrographs at higher redshifts ($z \gtrsim 0.5$, e.g. BPT diagrams or WHAN diagram), or 2) higher-order emission lines are very difficult to detect and demand high signal-to-noise ratios (S/N) for a good measurement (e.g. FLR or OHNO diagrams). Alternative optical classification schemes, such as the ones suggested by \cite{Rola97}, use bluer lines  (e.g. $[\text{OII}]\lambda\lambda3727,3729$ doublet) which avoids some of these issues, but attributing a class to galaxies with these diagrams is a non-trivial matter \citep{paalvast,perrotta}.

As we venture into `intermediate' redshifts ($1.5 < z < 2.5$), galaxy classification with optical spectrographs becomes a significant challenge, even considering all of these alternative schemes. `Illuminating the fog' over galaxies at these epochs is essential, as this is the most active period of the Universe when it comes to star formation \citep{madaudickinson_14}. A classification scheme in the optical that can cover these redshifts and not run into any of the aformentioned problems would allow us to better understand how galaxies change and evolve over time. 

In this paper, we aim to provide a potential scheme that could solve some of these issues, by re-exploring a known diagram, based on the approach of the WHAN diagram \citep{cidfernandes} and the philosophy of \cite{Rola97}, where we utilise, respectively, EWs on one axis and emission lines on the bluer end of the optical spectrum on the other. This diagram compares the EW of $\hbeta$ with the widely used $\oiii/\hbeta$ diagnostic. We call it the OB-I diagram. We are not the first to notice that this classification scheme is good at separating galaxy types (see Fig. 8 of \citealt{Teimoorinia18}), but a full exploration of this diagram in a wide range of epochs ($0 < z < 2.7$) is in order, including analysing the separate populations present in it and looking at it through theoretical models, to properly quantify its power as a galaxy classification tool.

This paper is structured as follows: in Sect. \ref{section:sample} we describe both the sample used as well as the selection criteria; in Sect. \ref{section:results} we describe the results of the spectral analysis of the low redshift galaxies; in Sect. \ref{section:discussion} we discuss these results; in Sect. \ref{section:highz} we apply the diagram to the higher-redshift sample and in Sect. \ref{section:conclusion} we provide our conclusions. Appendix \ref{appendix:highz class} provides a detailed discussion of how AGNs are selected in a higher redshift sample. Throughout this work, we use a cosmology with $H_0 = 70 \; \text{km s}^{-1} \text{Mpc}^{-1}, \Omega_M = 0.3$ and $\Omega_{\Lambda}=0.7$ and the \cite{Chabrier_2003} initial mass function (IMF).

\section{Sample}\label{section:sample}

In order to accurately study the classification of galaxies with the OB-I diagram across a wide range of redshifts, we decided to explore several spectroscopic datasets. In this section, we describe each spectroscopic survey and our selection for each of them.

\subsection{FADO-SDSS}
\label{subsect:fadosdss}

The Sloan Digital Sky Survey (SDSS, \citealt{sdss_technical1,sdss_technical2,sdss_technical3}), conducted with the 2.5m telescope at the Apache Point Observatory, mapped a quarter of the area of the sky, providing both photometry and spectroscopy. Our work uses the seventh Data Release\footnote{\texttt{\hyperlink{http://classic.sdss.org/dr7/}{http://classic.sdss.org/dr7/}}} of this survey (SDSS-DR7, \citealt{sdss}), that covered $9\,380$ deg$^2$ of the sky, where $929\,555$ galaxy spectra were measured through fibres with a sky aperture of 3 arcsec. The spectroscopic measurements covered wavelengths between $3\,800$ and $9\,200$Å, with resolution of $1\,800$ to $2\,200$ and signal-to-noise ratio, $\text{S/N}>4$ per pixel at $g$ band magnitude $g=20.2$.

In \cite{cardoso}, the SDSS-DR7 survey was processed in the following way: firstly, the spectra were corrected for galactic foreground extinction, using the \cite{Schlegel_1998} dust maps and applying the correcting factor of \cite{Schlafly_2011}. They also opted to use the reddening law defined in \cite{Cardelli_1989}. Secondly, they applied the population spectral synthesis code Fitting Analysis using Differential evolution Optimization (FADO, \citealt{fado}) with 150 \cite{Bruzual_2003} simple stellar populations, a \cite{Chabrier_2003} IMF and Padova 1994 evolutionary tracks \citep{Alongi_1993,Bressan_1993,Fagotto_1994a,Fagotto_1994b,Girardi_1996}. This stellar spectra library contains populations with 25 ages (between 1 Myr and 15 Gyr) and six metallicites ($1/200;1/50;1/5;2/5;1;2.5 \; Z_{\odot}$). The extinction law assumed was the \cite{calzetti_2000} law.

This process was applied to $926\,246$ galaxies. We used their sample and further selected galaxies that had flux measurements with $\text{S/N}>3$ for the emission lines of $\halpha$, $\hbeta$, $\nii$ and $\oiii$. This dropped our count to $161\,087$ galaxies. This sample has a median redshift and standard deviation of $\langle z \rangle = 0.07 \pm 0.05$. We call this the FADO-SDSS sample.

\subsection{LEGA-C}

The Large Early Galaxy Astrophysics Census (LEGA-C, \citealt{legac1,legac2}) is a public spectroscopic survey from the Visible Multi-Object Spectrograph (VIMOS, \citealt{vimos}) at the Very Large Telescope (VLT) in the European Southern Observatory (ESO). This survey covered 1.42 deg$^2$ of the sky, between redshifts of $0.3 < z < 1.0$, and produced $4\,081$ galaxy spectra, with nearly all objects possessing spectroscopic redshifts, velocity dispersion measurements, emission line fluxes and equivalent widths, as well as absorption line indices.

We focus on the final LEGA-C spectroscopic catalogue\footnote{\texttt{\hyperlink{https://users.ugent.be/$\sim$avdrwel/research.html}{https://users.ugent.be/$\sim$avdrwel/research.html}}}, described in detail in \cite{legac_cat}. From this catalogue, we selected galaxies with $\text{S/N}>3$ for the emission lines of $\hbeta$ and $\oiii$ (due to the redshift range, $\halpha$ and $\nii$ are not available in the spectrograph). Because some EWs have negative values, we ensured that only positive values of the EW of $\hbeta$ were selected. Furthermore, we also removed all the galaxies that had \texttt{flag\_spec}$\,= 2$, as these spectra had problems in the flux calibration process. This selections drops the count to $723$ galaxies. This sample has a median redshift and standard deviation of $\langle z \rangle = 0.68 \pm 0.07$.

\subsection{VANDELS}

VANDELS\footnote{\texttt{\hyperlink{http://vandels.inaf.it/}{http://vandels.inaf.it/}}} is a deep VIMOS survey on the Chandra Deep-Field South (CDFS, \citealt{cdfs}) and Ultra Deep Survey (UDS) fields \citep{vandels1,vandels2} conducted at the ESO/VLT. In its latest data release \citep{vandels_cat}, a total of $2\,165$ spectra were measured, in the wavelength range of $4\,800 - 9\,800$Å, with a spectral resolution of $R \simeq 250$ and a mean dispersion of $2.5$Å/pixel.

From the spectroscopic catalogue of the latest data release, we selected galaxies with $\text{S/N}>3$ in the emission lines of $\hbeta$ and $\oiii$, specifically in the CDFS field. There are no galaxies in the UDS field in VANDELS with $\text{S/N}>3$ for these lines. The $\halpha$ and $\nii$ emission lines do not meet the $\text{S/N}>3$ criterion, so we do not use these values from this survey. We also checked the flagging system and found that our sample had no AGNs flagged. This leaves our sample with a total of $3$ galaxies, all with redshifts between $0.4 < z < 0.75$.

\subsection{3D-HST}

The 3D-HST survey\footnote{\texttt{\hyperlink{https://archive.stsci.edu/prepds/3d-hst/}{https://archive.stsci.edu/prepds/3d-hst/}}} \citep{brammer_12} is a near-infrared spectroscopic survey realised by the Hubble Space Telescope, over four fields: AEGIS (All-Wavelength Extended Groth Strip International Survey, \citealt{aegis}), COSMOS (Cosmological Evolution Survey, \citealt{cosmos}), GOODS (Great Observatories Origins Deep Survey, \citealt{goods-s}) and UDS. This catalogue possesses grism spectra for $98\,668$ individual galaxies \citep{momcheva_16}, which have redshifts, emission line fluxes and equivalent widths. Furthermore, each field contains photometric measurements from many instruments \citep{skelton14}, but we are interested in the data from the Infrared Array Camera (IRAC, \citealt{irac}) aboard the Spitzer space telescope \citep{Spitzer}, in order to compare the OB-I classification with infrared classification. 

From the grism spectra data, we selected galaxies with $\text{S/N}>3$ for the emission lines of $\hbeta$ and $\oiii$, ensuring that the equivalent width of $\hbeta$ had a positive measurement. The $\halpha$ and $\nii$ emission lines are convolved together, so we decided not to use them. This selection drops the count to $945$ galaxies, with a median redshift and standard deviation of $\langle z \rangle = 1.8 \pm 0.3$. 

When it comes to the IRAC photometric measurements, we used the photometric catalogue available and cross-matched the sky positions with the spectroscopic catalogue. All galaxies in our selection had matches in the photometric catalogue within a 1 arcsecond radius. To ensure that the IRAC measurements match the same criteria as the spectroscopic catalogue, we enforced $\text{S/N}>3$ on all the IRAC flux densities, when they were available. These measurements are $F_{3.6 \mu m}$, $F_{4.5 \mu m}$, $F_{5.8 \mu m}$ and $F_{8.0 \mu m}$, and represent flux densities at 3.6, 4.5, 5.8 and 8.0 microns. This left us with a total spectroscopic-photometric crossmatch sample of 346 galaxies.

\subsection{MOSDEF}

The MOSFIRE (Multi-Object Spectrometer for Infra-Red Exploration, \citealt{mosfire}) instrument produced the MOSDEF (MOSFIRE Deep Evolution Field, \citealt{kriek_15,reddy_15}) survey. This survey selected galaxies to observe in three redshift ranges: $1.37 \leqslant z \leqslant 1.80$;  $2.09 \leqslant z \leqslant 2.61$; and $2.95 \leqslant z \leqslant 3.80$. These ranges were selected so that bright emission lines (e.g. $\halpha, \hbeta$, see Fig.1 of \citealt{kriek_15}) fall within the low transmission windows of Earth's atmosphere in the near-infrared. 

We make use of the emission line catalogue present in the MOSFIRE website\footnote{\texttt{\hyperlink{https://mosdef.astro.berkeley.edu/for-scientists/}{https://mosdef.astro.berkeley.edu/for-scientists}}}, for a total of $1\,824$ galaxies. We selected galaxies with $\text{S/N}>3$ for the emission lines of $\halpha$, $\hbeta$, $\nii$ and $\oiii$, ensuring that $\hbeta$ and $\halpha$ had measurable EWs. With this selection, this drops the count to 70 galaxies, distributed in two redshift ranges: $1.4 < z < 1.7$ and $2.1 < z < 2.6$.

\subsection{FADO-JWST}

The James Webb Space Telescope (JWST, \citealt{jwst}) has many surveys to its name. We opted to use data from the JWST Advanced Deep Extragalactic Survey\footnote{\texttt{\hyperlink{https://archive.stsci.edu/hlsp/jades}{https://archive.stsci.edu/hlsp/jades}}} (JADES, \citealt{jades1,jades5,jades2,jades3,jades4}). This survey was exploited by Miranda et al. (in prep), which used the FADO spectral fitting tool to quantify the impact of the nebular component in higher redshift galaxies.

Before applying FADO to the spectra, Miranda et al. (in prep) de-redshifted and rebinned the spectra to a step of 1 Å and corrected for galactic extinction considering the \cite{Schlegel_1998} dust maps and the extinction curve from \cite{Cardelli_1989} with $R_V=3.1$. Afterwards, the authors considered the spectral range between $3\,000 - 9\,000$ Å and a spectral basis composed of simple stellar populations from \cite{Bruzual_2003}, considering a \cite{Chabrier_2003} IMF and Padova 1994 evolutionary tracks. The main basis includes 57 ages, between 0.5 Myr and 13 Gyr, and 3 metallicities, $Z=0.2,0.4$ and $1 \; Z_{\odot}$. The set of simple stellar populations used for each galaxy is obtained by removing from the main basis the simple stellar populations that are older than the age of the Universe at the redshift of the galaxy.

This process was applied to ten SF galaxies in the JADES survey. With the $\text{S/N}>3$ condition on $\hbeta$ and $\oiii$, all of these objects are kept, and all have measurable EWs. These ten galaxies have a wide range of redshifts, varying from $0.43$ to $2.71$. We named this the FADO-JWST sample. 

\vskip 0.3cm

\begin{figure}[h]
    \centering
    \includegraphics[width=\linewidth]{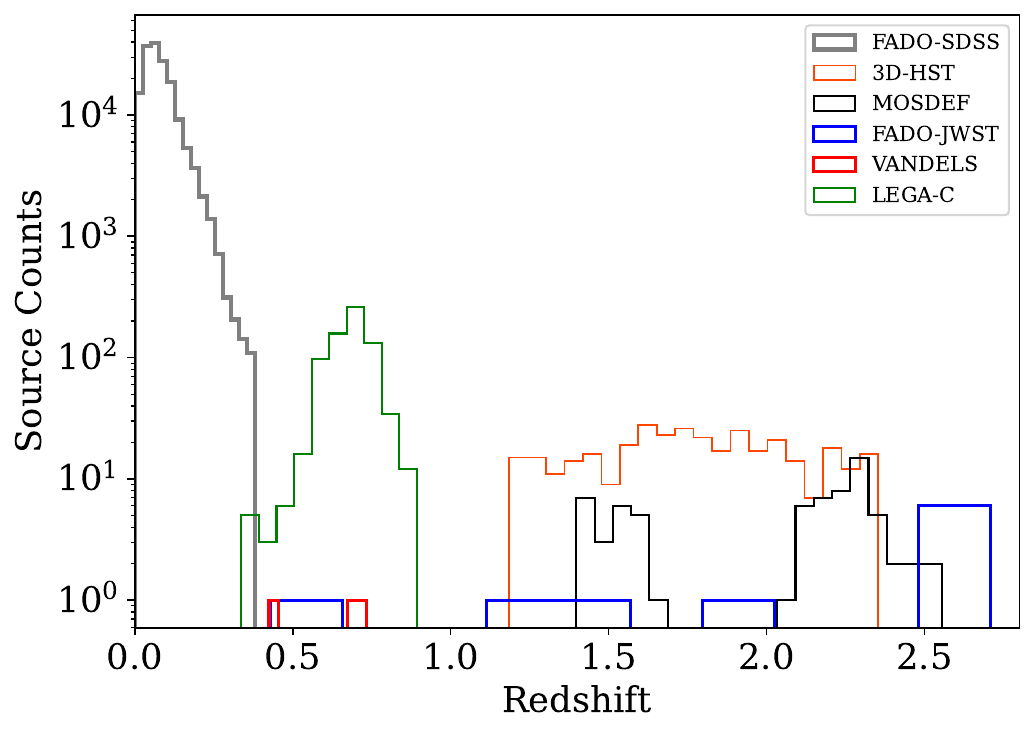}
    \caption{Redshift distribution of the galaxies in all spectroscopic datasets.}
    \label{fig:all_redshift}
\end{figure}

The information from all these catalogues and our selection is summed up in the redshift distribution in Fig. \ref{fig:all_redshift} and the total count of galaxies is located in Table \ref{tab:selection}.

\begin{table}[h!]
\caption{Number of galaxies per spectroscopic dataset, with conditions as defined in the text.}
\centering
\label{tab:selection}
\begin{tabular}{llccc}
\multicolumn{1}{c}{Dataset} & \multicolumn{1}{c}{Source Counts} & Range & \multicolumn{1}{c}{$\langle z \rangle$}  \\ \hline
FADO-SDSS & 161 087 & $0 < z < 0.4$    & 0.07 \\
LEGA-C    & 723     & $0.3 < z < 0.9$  & 0.68 \\
VANDELS   & 3       & $0.4 < z < 0.8$  & 0.61 \\ 
3D-HST    & 346     & $1.2 < z < 2.4$  & 1.8 \\
MOSDEF    & 70      & $1.4 < z < 2.6$  & 2.0 \\
FADO-JWST & 10      & $0.4 < z < 2.7$ & 2.0 \\ \hline
Total     & 162 239 & $0 < z < 2.7$    & 0.1 \\ \hline
\end{tabular}
\end{table}

\section{Low-redshift results}\label{section:results}

In this section, we re-explore a diagram with blue emission lines, using both a combination of EWs and flux ratios to understand the properties of galaxies.

Emission line ratios measure the properties of the gas inside a galaxy (i.e. gas density, ionisation parameter). EWs, instead, measure the properties of the ionisation inside the galaxy (i.e. number of ionising photons, power of the ionising agent), assuming the fraction of escaping ionising photons is negligible \citep{cidfernandes_2010,stasinka_2018}.

\begin{figure*}[h]
    \centering
    \includegraphics[width=\linewidth]{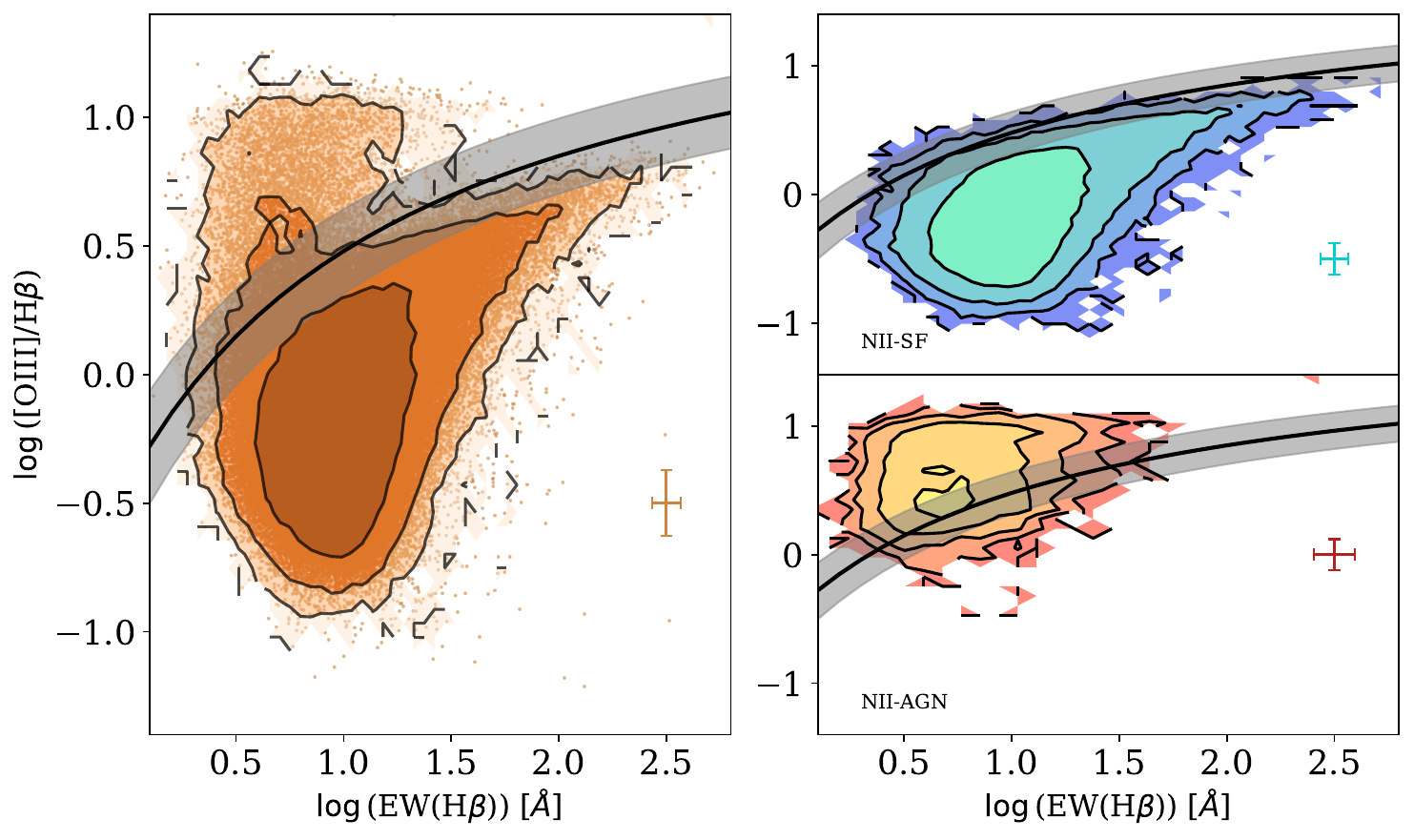}
    \caption{The OB-I diagram, with our empirical separation line. On all plots, the black curve and the grey shaded region represent our empirical separation line, seen in Eq. \ref{eq:hbeta limit}. \textit{Left:} The orange dots represent the FADO-SDSS sample as described in Sect. \ref{subsect:fadosdss}. The error bar represents the median error for all the points. Each contour represents 25\% more of the sample, from the smallest to the largest. \textit{Right:} The top panel represents the NII-SF galaxies, while the bottom panel represents the NII-AGN galaxies, which are what we used to define the empirical line.}
    \label{fig:obi_separation}
\end{figure*}

Combining both of these observables should be a valuable tool to help us separate galaxy types. There are many examples of this approach, but one that is rarely used to separate galaxies comes in the form of comparing the EW of $\hbeta$ with the emission line flux ratio of $\oiii$ and $\hbeta$. We named it the OB-I (read as Oh-Bee-One) diagram, as it uses Ionised states of Oxygen and hydrogen (namely, hydrogen Beta).

The FADO-SDSS sample (described in Sect. \ref{subsect:fadosdss}) provides us with a sample of a high volume of galaxies in the Local Universe and beyond ($z \lesssim 0.4$), where the classification capabilities of the OB-I can be tested. The distribution of this data in the OB-I diagram can be seen in the left panel of Fig. \ref{fig:obi_separation}. Looking at this figure, we can see that there seems to be two main clusters of data in the parameter space: a `cloud' on the top that has more dispersed data, and a `teardrop' region which is more densely populated.

To separate these two regions, we can proceed through three main ways: using fully theoretical models (e.g. \citealt{kewley_2001}), fully empirical models (e.g. \citealt{veilleux}) or a combination of both, in a semi-empirical model (e.g. \citealt{kauffmann_2003}). Since we are basing ourselves on the results from the data alone, we opted for a fully empirical approach. However, using only the two apparent clusters as a basis of galaxy classification is unclear, as the edges of a distribution are bound to be less dense and possess more scatter. As such, we used the classification from the NII diagram (or comparing $[\text{NII}]\lambda6584/\text{H}\alpha$ with $[\text{OIII}]\lambda5007/\text{H}\beta$) as a basis to separate these two apparent galaxy populations. There are three types of galaxies, using the separation lines from \citet{kewley_2001} and \citet{kauffmann_2003}: SF, Composite and AGN, which from now on we call NII-SF, NII-Composite and NII-AGN. We have $137\,181$ NII-SF galaxies, $16\,990$ NII-Composites and $6\,890$ NII-AGNs in the FADO-SDSS sample. We focus on the NII-SF and NII-AGN to create our separation line.

\begin{figure}[h]
    \centering
    \includegraphics[width=1.0\linewidth]{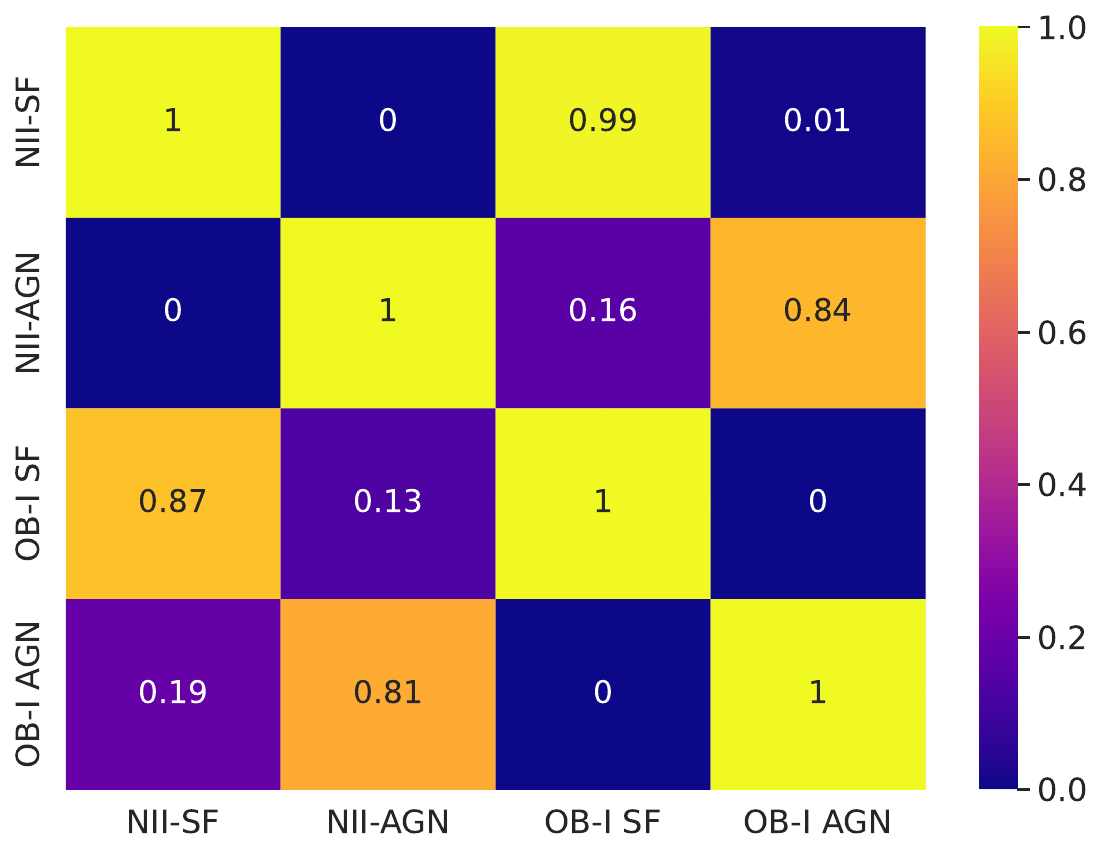}
    \caption{Match between NII and OB-I classes. This tells us how many galaxies in one classification belong to another, e.g., if we look at the bottom left square, we can see that $19\%$ of OB-I AGNs are classified as SF by the NII diagram.}
    \label{fig:nii obi percentage}
\end{figure}

From the right panel of Fig. \ref{fig:obi_separation}, we can see that the NII-AGNs are mostly concentrated on the top left of the OB-I diagram, generally occupying a different space from the NII-SF galaxies, which are spread throughout the parameter space, preserving the `teardrop' shape that was mentioned previously. In order to separate these two now evident populations -- SF and AGN -- we focused on fitting a general hyperbolic line with three parameters, given by the following equation:

\begin{equation}
    \log\left( \frac{\oiii}{\hbeta} \right) = \frac{a}{\log(\text{EW}(\hbeta)) + b}+c
     \label{eq:hyperbole}
\end{equation}
We decided on this equation form due to the fact that the frontier of the NII-SF population has a hyperbolic shape. To fit the three parameters on the equation, we enforced a simple criterion: the number of NII-SF and NII-AGNs that cross our empirical separation line has to be minimised, that is, we want the smallest amount of NII-SF galaxies crossing into the AGN region, and vice-versa. Calculating the variables of the hyperbole by brute-force estimation, we found that the best equation that fits this criteria is the following:
\begin{multline}
    \log\left( \frac{\oiii}{\hbeta} \right) = \frac{-3.00}{\log(\text{EW}(\hbeta)) + (1.39\pm0.07)} \\ + (1.73\pm0.13)
    \label{eq:hbeta limit}
\end{multline}
The uncertainties associated with our values come from the median uncertainties associated to all the galaxies in our sample. Overall, we find that 521 NII-SF galaxies (or $13\%$ of the full NII-SF population) cross over into the AGN region, while $1\,068$ NII-AGNs (or $18\%$ of the full NII-AGN population) cross over into the SF region. In Fig. \ref{fig:nii obi percentage}, we can see these results, as well as when we adopt our current empirical classification and apply them to the NII diagram (in Sect. \ref{subsec:compare} we provide more insights).

Something to note is that the EWs measured by FADO are about 25\% higher than the EWs measured using the pseudo-continuum \citep{henrique}. We checked to see if the position of our separation line was dependent on the methodology used to measure EWs: by adopting the measurements provided in the MPA-JHU analysis for the same galaxies\footnote{\texttt{\hyperlink{https://wwwmpa.mpa-garching.mpg.de/SDSS/DR7/}{https://wwwmpa.mpa-garching.mpg.de/SDSS/DR7/}}} \citep{brinchmann_2008}, who used a different measurement method from the one used in FADO, resulted in very similar positions of both the NII-SF and NII-AGN populations (a maximum of 0.1 dex separation). This allows us to conclude that our separation line remains valid even for different ways of measuring EWs.

We are now in possession of a simple optical classification scheme, which divides the FADO-SDSS sample in two: $153\,983$ objects are SF galaxies and $7\,104$  are AGNs (considering the line only, and not the uncertainties). In order to further understand what drives the separation between these two galaxy populations, we can look at the emission line ratio and the EW and try to understand the physical processes at play. 

High values of the emission line flux ratio of $\oiii$ and $\hbeta$ primarily reflects a high ionisation parameter (see for example Fig. 5 of \citealt{brinchmann_23}) or can be reproduced by shocks \citep{allen08}. In the case of the former, higher values of the ionisation parameter can occur typically either in low-metallicity starburst galaxies (like Blue Compact Dwarfs) or in AGN. As for shocks, these usually require galaxies interacting/merging. Statistically, we don't expect our FADO-SDSS sample, which is a low redshift sample, to be dominated by low metallicity starburst galaxies nor by mergers/interactions between galaxies, so the upper part of the OB-I diagram should likely be populated by AGNs.

When it comes to the EW of $\hbeta$, a combination of different physical mechanisms are at play, complicating the interpretation of the values observed. On one hand, the EW in SF galaxies scales linearly with the specific star-formation rate of a galaxy (defined as the star-formation divided by the stellar mass of the galaxy, see Fig. 3 of \citealt{casado15}), implying that high star-formation rates gives us higher equivalent widths. On the other hand, the EW also depends on numerous other factors, such as gas-phase metallicity \citep{papaderos23}, Lyman-continuum escape fraction \citep{papaderos2013} and we cannot discount the effects of the aperture of the SDSS fibres \citep{patricio}. Due to the fact that AGNs have a higher production rate of the Lyman-continuum than SF galaxies, one would expect them to have higher EWs, but the AGN continuum is both thermal and non-thermal, making it very bright, which in turn reduces the EW of an emission line.

Combining both the information from the flux ratio and EWs, it makes sense that AGNs -  which are objects with high ionisation parameters, harder ionising fields, typically low star-formation rates and high continua values - are `trapped' in the upper left corner of the diagram, where the $\oiii/\hbeta$ flux ratio is high and EWs are low, while SF galaxies - which have a higher degree of star-formation and usually lower continua than AGNs - are mostly in the lower parts of the diagram and can reach significantly higher EWs.

\section{Discussion}\label{section:discussion}

In the previous section, we created a simple empirical line in the OB-I diagram that is expected to separate galaxies into two types: SF-dominated galaxies and AGNs. In this section, we dive deeper into the ramifications of our classification: first, we take into accounts some caveats in the NII diagram classification; secondly, we compare the OB-I classification with another BPT diagram; and finally, we test our empirical line against theoretical models, to find the physical origin for this separation.

\subsection{Notes on the NII diagram}\label{subsec:compare}

The NII diagram is one of the most commonly used classification schemes, having been studied by several authors to distinctly separate galaxy types (e.g. \citealt{bpt,veilleux,kewley_2001,kewley,kauffmann_2003,schaw}). However, it is known that the NII diagram has some difficulties classifying galaxies with low stellar mass and, more importantly, sub-solar metallicity, due to the relative decrease of nitrogen emission in comparison with hydrogen emission \citep{groves06}, placing these objects in the SF region when they really are very likely to possess an AGN \citep{kewley_2013,polimera22,harish23}.

\begin{figure}[h]
    \centering
    \includegraphics[width=1.0\linewidth]{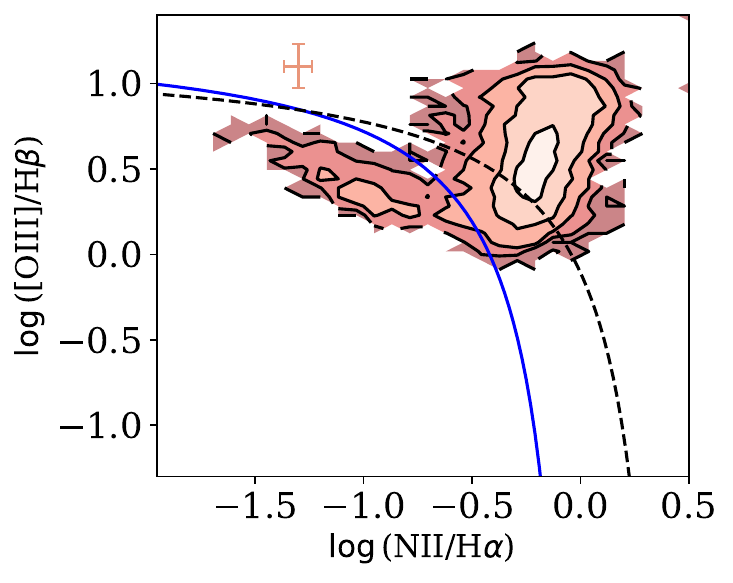}
    \caption{OB-I AGNs in the NII diagram. The blue line represents the \citet{kauffmann_2003} separation line, while the black dashed line represents the \cite{kewley_2001} line. All remaining elements represent the same as in Fig. \ref{fig:obi_separation}.}
    \label{fig:obi agns bpt}
\end{figure}

In Fig. \ref{fig:obi agns bpt}, looking at the OB-I AGNs in the NII diagram, we can see that there is a separate population of these objects present in the SF region of the NII diagram. In order to see if the some of the discrepancies are explained by the effects of sub-solar metallicity, we made use of the MPA-JHU measurements of metallicity of SDSS-DR7 \citep{brinchmann,Tremonti_2004}, as the FADO analysis does not provide nebular metallicities. This gives us metallicity measurements for $548$ of the OB-I AGNs that reside in the NII-SF or NII-Composite regions, out of which $90\%$ have sub-solar metallicities. Due to this, we removed these galaxies from our analysis, though their true nature is likely to be an AGN rather than a SF galaxy.

\begin{figure}[h]
    \centering
    \includegraphics[width=1.0\linewidth]{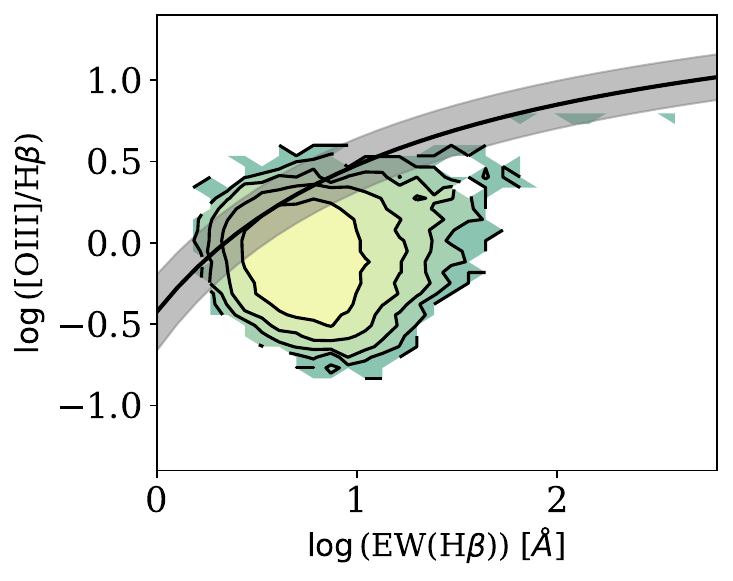}
    \caption{NII-Composites in the OB-I diagram. All remaining elements represent the same as in Fig. \ref{fig:obi_separation}.}
    \label{fig:composite obi}
\end{figure}

\begin{figure*}
    \centering
    \includegraphics[width=1.0\linewidth]{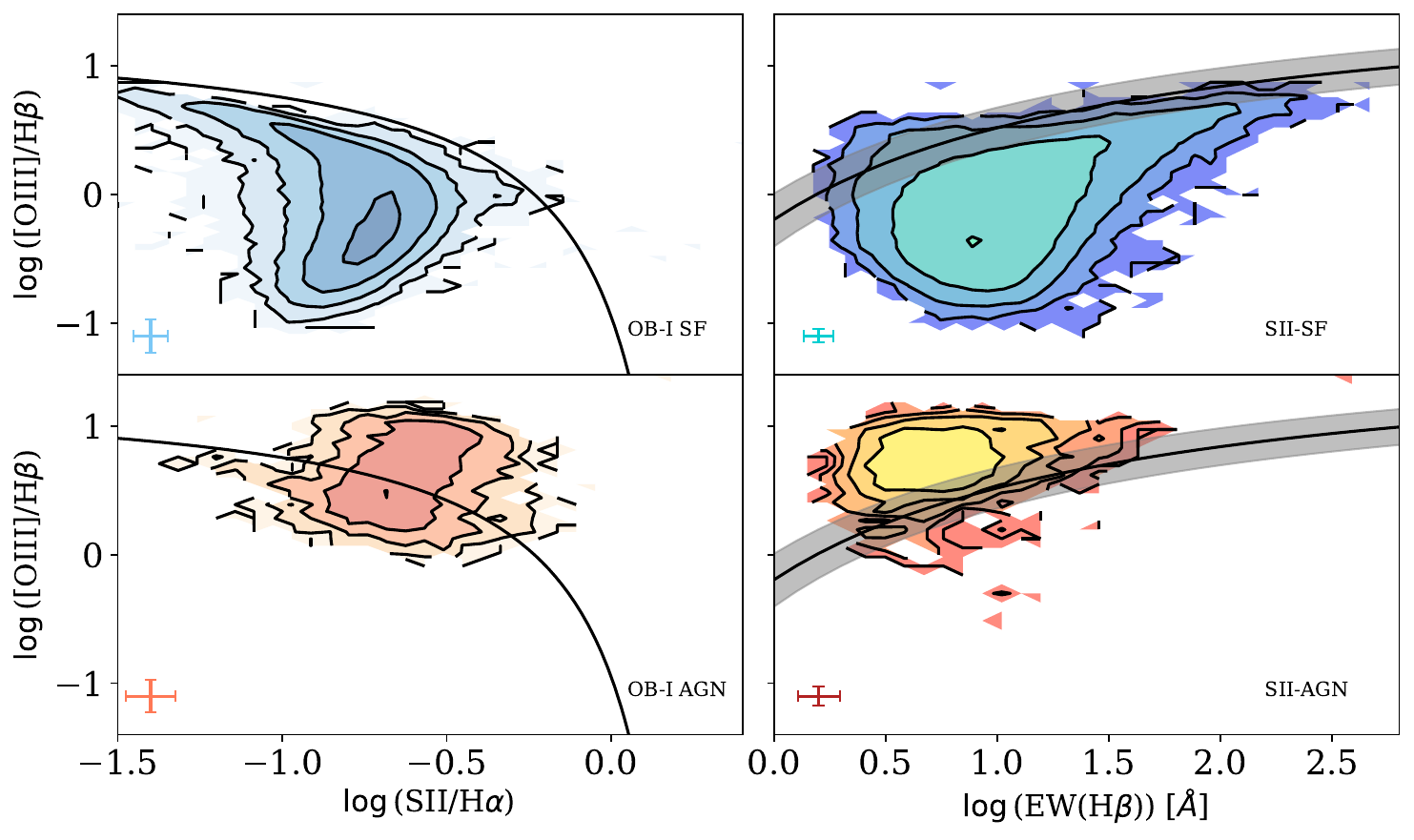}
    \caption{Comparison between the SII diagram (left) and the OB-I diagram (right). In the OB-I diagram, the black line and the shaded region represent Eq. \ref{eq:hbeta limit}. In the SII diagram, the solid black line represents the separation by \cite{kewley_2001}. Each contour represents 20\% more of the sample and the error bars represent the median error for each axis, for each classification. On the left, we have the OB-I classifications on the SII diagram; on the right, we have the SII classifications on the OB-I diagram. Top panels represent SF galaxies and bottom panels represent AGNs.}
    \label{fig:bpt obi}
\end{figure*}

When it comes to the NII-Composites, a direct comparison between the NII and OB-I diagrams is not clear. Firstly, this classification arises as a mixture between the semi-empirical fit of \cite{kauffmann_2003} and the theoretical fit of \cite{kewley_2001} in the NII diagram. Since they are not pure AGNs nor simply SF galaxies, they have been joined together as a type of galaxy whose emission comes from both sources. However, we can clearly see from Fig. \ref{fig:composite obi} that these objects occupy a clear region in the parameter space, distinct from the AGN region and in the core of the SF region. Exploring these galaxies with more detail demands the use and comparison with a theoretical photoionisation model, which is what we will do in Sect. \ref{subsec:models}.

\subsection{Comparison with SII diagram}\label{section:compare}

Comparing the OB-I diagram with other classification schemes of the same nature allows us to have a better insight into how this diagram separates galaxy types. We have already used the NII diagram as a starting point to create our empirical line, but it is interesting to compare it with a different BPT diagram, since they provide different information on the same set of galaxies. As such, in this section we focus on the SII diagram (which compares the $\sii/\halpha$ and $\oiii/\hbeta$ ratios).

\begin{figure}[h]
    \centering
    \includegraphics[width=1.0\linewidth]{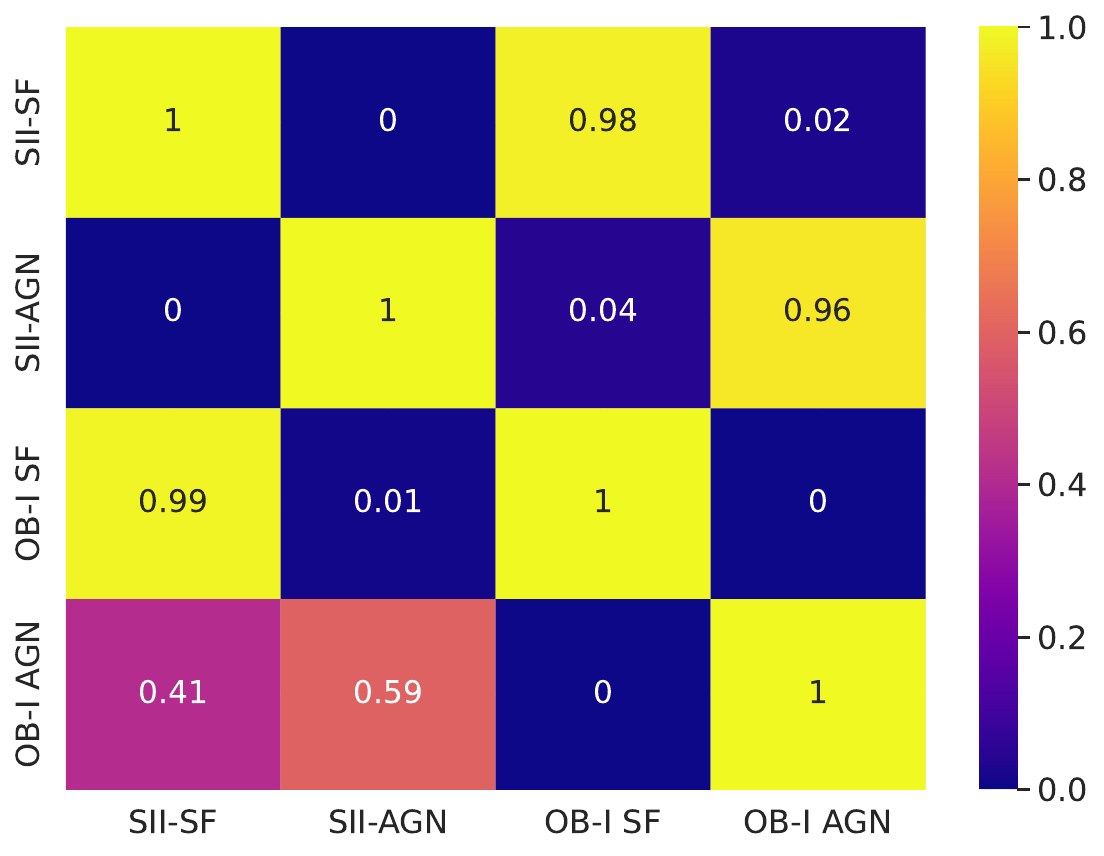}
    \caption{Match between SII and OB-I classes. All elements represent the same as in Fig. \ref{fig:nii obi percentage}.}
    \label{fig:sii obi percentage}
\end{figure}

\begin{figure*}[h]
    \centering
    \includegraphics[width=\linewidth]{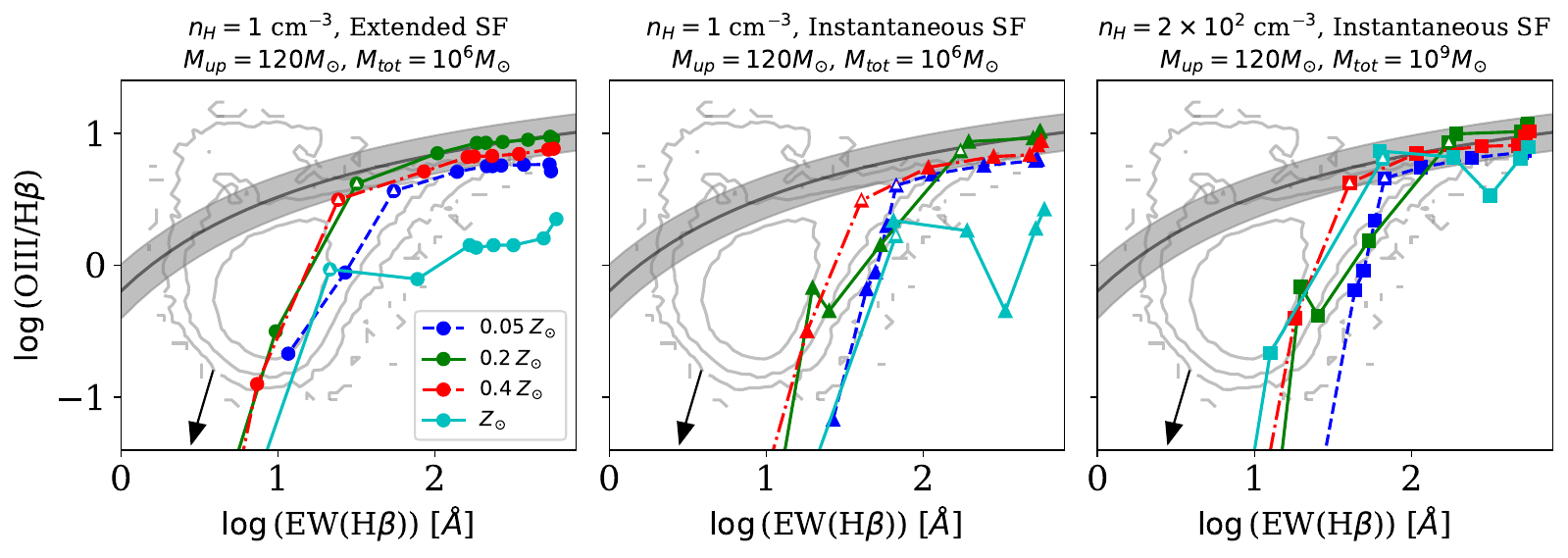}
    \caption{Evolutionary models from \cite{stasinska01} overplotted on the OB-I diagram. From left to right, we have scenarios 1), 2) and 3) as described in the text (see Sect. \ref{subsec:models}). For all plots, the gray contours represent the FADO-SDSS sample; the black line our separation line given by Eq. \ref{eq:hbeta limit}; the black arrow shows how the star-forming regions evolve with time; the blue markers and dashed line represent the model with $0.05 \: Z_{\odot}$, the green markers and full line represent the model with $0.2 \: Z_{\odot}$; the red markers and dot-dashed line the model with $0.4 \: Z_{\odot}$; and the cyan markers and full line the model with $Z_{\odot}$. The first marker of each metallicity is always on the top right of the OB-I diagram, and each subsequent marker represents a 2 Myr step for scenario 1) and a 1 Myr step for scenarios 2) and 3). The white triangles inside some markers represent the point where the massive stars fade: 14 Myrs for scenario 1) and 5 Myrs for scenarios 2) and 3).}
    \label{fig:models}
\end{figure*}

We distinguish galaxy types in the SII diagram by using the \cite{kewley_2001} demarcation line, where we are presented with two classes: SII-SF and SII-AGN. In Fig. \ref{fig:bpt obi}, we compare the classifications of the SII diagram with the OB-I diagram. Overall, we find a good agreement between these diagrams, where $98\%$ of the SII-SF are correctly classified in the OB-I diagram, and $96\%$ of the SII-AGN are correctly classified as well. On the other side, if we take the OB-I classification and apply it to the SII diagram, we also find a remarkable agreement with the SII classification, where $99\%$ of the OB-I SF galaxies are in the SF region, and $59\%$ of the OB-I AGNs are in the AGN region. All of this information is summed in Fig. \ref{fig:sii obi percentage}.

When comparing the results of the NII and SII diagrams (Figs. \ref{fig:nii obi percentage} and \ref{fig:sii obi percentage}), we expect the NII to have higher values of matched classification with the OB-I diagram rather than the SII diagram, as our empirical classification is based on the NII diagram classes. Although that is true from the OB-I classification side, if we focus on the BPT diagram classes, we find that, surprisingly, the SII diagram has higher values of matched AGNs when compared with the NII diagram. This is likely due to two reasons: 1) the SII diagram is more sensitive to `weaker' AGNs \citep{polimera22}, and the OB-I diagram is likely tracing these objects as well and 2) the NII diagram possesses the Composite objects, which are not distinguishable in either the SII or OB-I diagrams.

\subsection{Theoretical Models}\label{subsec:models}

We apply the evolutionary theoretical models of HII regions in SF galaxies, computed by \cite{stasinska01}, onto the OB-I diagram, in order to understand what the theory says about our empirical separation. These models all start at $10^4$ years, and follow a timestep of $10^6$ years and assume a \cite{salpeter} IMF. We chose to compare three scenarios: 1) an extended burst of star-formation (10 Myr) with an initial burst of $10^6 M_{\odot}$, upper stellar limit of $120 M_{\odot}$ in the IMF and gas density of $n_H = 1 \: \text{cm}^{-3}$; 2) an instantaneous period of star-formation with the same conditions as the previous scenario; and 3) an instantaneous period of star-formation, with an initial burst of $10^9 M_{\odot}$, upper stellar limit of $120 M_{\odot}$ in the IMF and gas density of $n_H = 2 \times 10^2 \: \text{cm}^{-3}$. For all these scenarios, we selected 4 metallicites: $0.05, 0.2, 0.4, 1 \: Z_{\odot}$. We chose these scenarios because they encompass the conditions of a wide variety of environments and star-formation activity across the period of a galaxy's lifetime, meaning we could probe the OB-I diagram for many kinds of galaxies, and not just one specific situation. We do not presume that these theoretical models represent all possible scenarios present in galaxies, but by looking at some extreme cases, we can still meaningfully infer the physics behind this diagram.

\begin{figure}[h!]
    \centering
    \includegraphics[width=\linewidth]{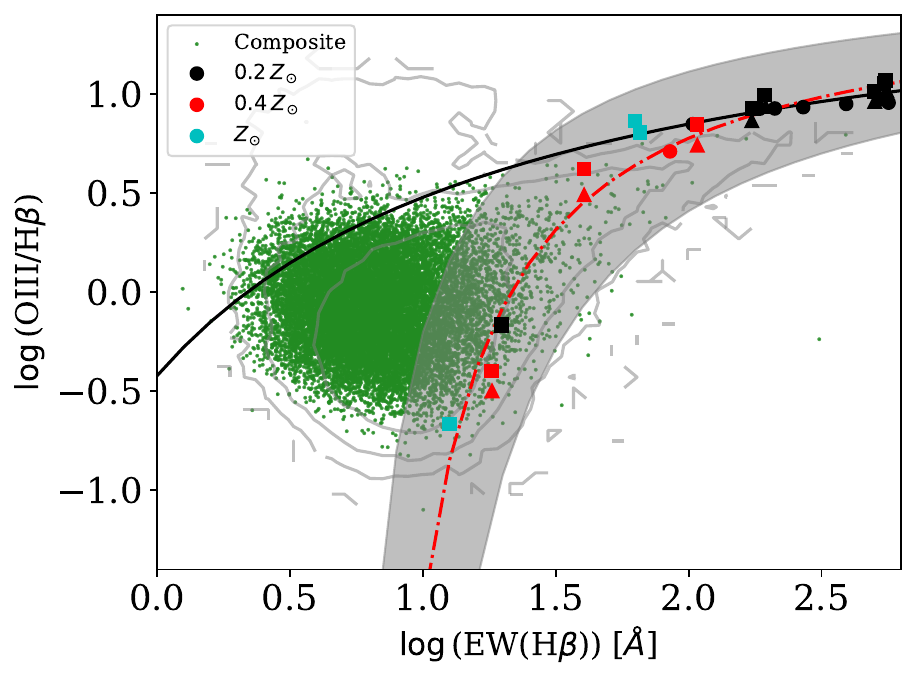}
    \caption{Best fit model for the third population in the OB-I diagram. The grey contours represent the FADO-SDSS sample; the green points the NII-Composite galaxies; the black line our separation line given by Eq. \ref{eq:hbeta limit}; the black symbols represent the $0.2 \: Z_{\odot}$ model for all scenarios; the red symbols represent the $0.4 \: Z_{\odot}$ model for all scenarios; the cyan symbols represent the $\: Z_{\odot}$ model for all scenarios. The red dot-dashed line represents the best fit for all the points in the diagram, given by Eq. \ref{eq:obi population limit}. The grey shaded area represents the $1\sigma$ uncertainty associated with the fit.}
    \label{fig:model fit}
\end{figure}

\begin{figure*}[h]
    \centering
    \includegraphics[width=1.0\linewidth]{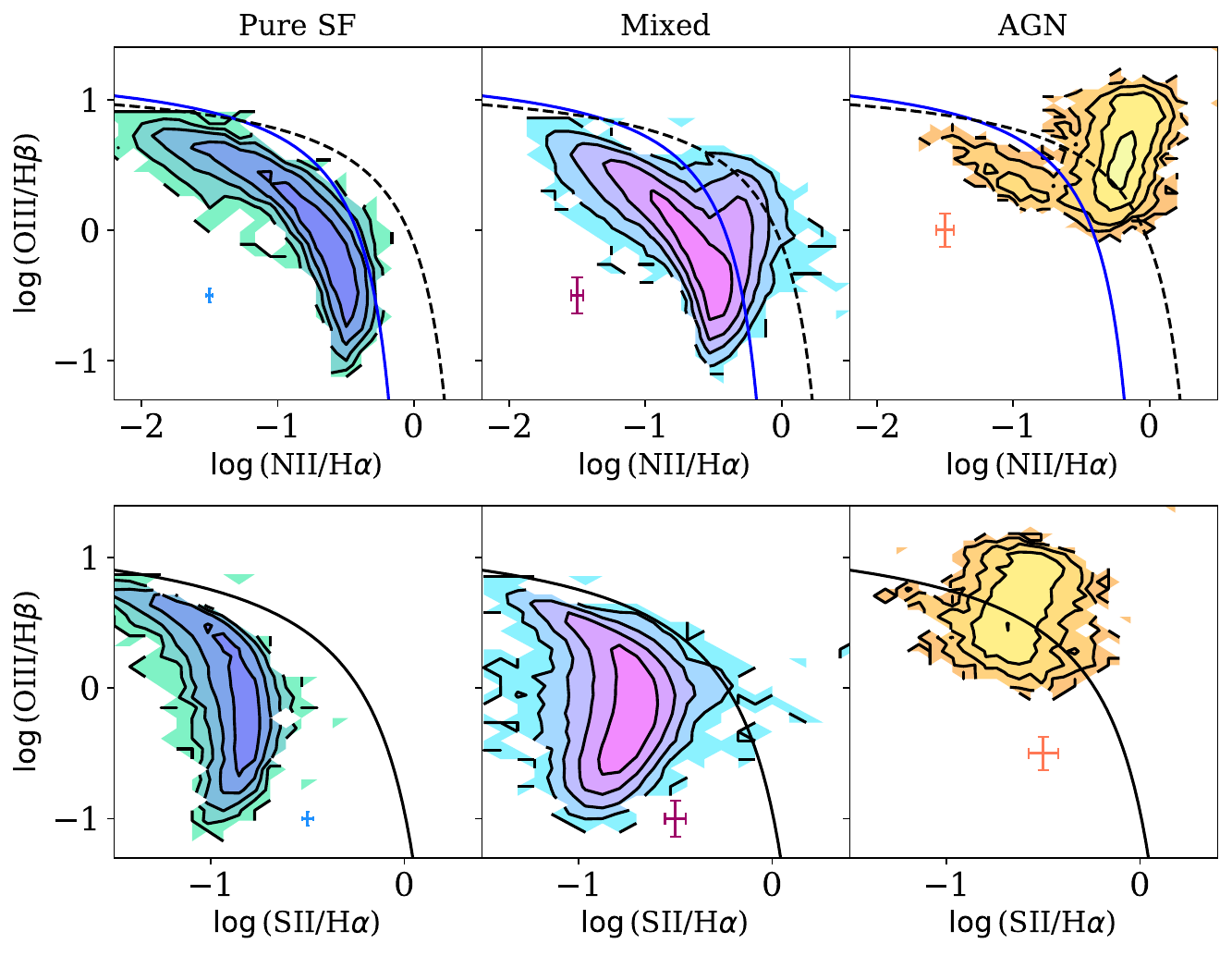}
    \caption{NII (top) and SII (bottom) diagrams, compared with OB-I diagram classification. \textit{Left:} The contours represent OB-I pure SF galaxies, where each contour represents 20\% more of the sample, and the error bar represents the median error for each axis. \textit{Middle:} The same as previously, but for OB-I mixed population. \textit{Right:} The same as previously, but for OB-I AGNs. All remaining elements represent the same as in Figs. \ref{fig:obi agns bpt} and \ref{fig:bpt obi}.}
    \label{fig:obi class bpt}
\end{figure*}

In Fig. \ref{fig:models}, the results of these computational models can be seen overplotted on the OB-I diagram. Across all scenarios, we can see that there is a downwards trend with time in the values of the axis of the OB-I diagram, signalled by the black arrow. In other words, the older a galaxy gets, the EW of $\hbeta$ and the $\oiii/\hbeta$ emission line ratio decline sharply. This sharp decline varies from scenario to scenario, starting at 14 Myr in scenario 1) and at 5 Myr in scenarios 2) and 3), but the meaning is the same: this represents when the massive stars fade, leaving only the less massive and less ionising stars, which in turn reduces the strength of the $\hbeta$ emission line relative to the $\oiii$ emission line \citep{byler17}.

Interestingly, we can see that, in the early stages of galaxies in all the scenarios, the $\oiii/\hbeta$ emission line ratio does not increase as the metallicity increases. There seems to be a distribution with a metallicity peak, meaning that for two different metallicities we obtain the same value for the $\oiii$ and $\hbeta$ ratio. This metallicity `degeneracy' has been noticed before in observations \citep{maiolino08,curti17,curti20} and more recent work confirms this effect (see Fig. 4 of \citealt{nakajima22}), where the values for the ratio peak at around $\approx 0.2 \: Z_{\odot}$, which is the same prediction as the models we are using.

Considering the $0.2 \: Z_{\odot}$ line as the edge limit case in all three scenarios, it is clear to see that for high values of the EW of $\hbeta$ (EW$(\hbeta) > 10^2$ Å) there is a very good agreement with our separation line in the OB-I diagram, indicating that this is the limit that a SF galaxy can reach. Beyond this limit, any further contribution must come from another emission type rather than star-formation, usually assumed to be an AGN. As we decrease the EW, especially in the scenarios where the SF episode is instantaneous, no model agrees with our separation line. In scenario 1), where there's an extended period of star-formation, the $0.2 \: Z_{\odot}$ model agrees with our line until EW$(\hbeta) \approx 10^{1.4}$ Å, but then we still have the aforementioned sharp decline. 

This drop is present in all models and scenarios we considered. If we look at the `core' of the SF region, where most galaxies lie, we find that these objects lie typically above this drop. Since the models we are working with only consider star-formation, then galaxies that have relatively high $\oiii/\hbeta$ emission line ratios and low EWs of $\hbeta$ likely possess a second process happening inside them that boosts the former but not the latter. In section \ref{section:results}, we already discussed that AGNs have high ionisation but retain low EWs, therefore, any object that is above the models and below our empirical line can belong to a mixed population, one that involves galaxies that are purely SF, possess AGNs, and have activity from both (akin to the NII-Composites).

In order to separate this potential population, which we called mixed population due to their nature, from the AGNs and galaxies that only possess star-formation, we selected the points from each scenario that were as close to our empirical separation line as possible for all ages, no matter the metallicity: by this we mean we selected the points that had the highest value of the $\oiii/\hbeta$ emission line ratio at each age. Looking at Fig. \ref{fig:model fit}, we can see that the mixed population is analogous to the NII-Composite galaxies, as they lie in the region above the points we chose. There are NII-SF galaxies in this region, but we argue that anything below our selection from the models should contain galaxies with star-formation as the main (or even exclusive) source of emission, without AGN contamination, as they are fully encompassed by the \cite{stasinska01} models - we call these sources pure SF galaxies.

Afterwards, we chose four types of functions to fit onto the points: a linear fit; a second degree polynomial; a third degree polynomial; and a hyperbolic function of the same type as Eq. \ref{eq:hyperbole}. To choose a function, we demand two things: firstly, that the chosen fit has the lowest $1\sigma$ errors and, secondly, that it includes as few NII-Composites as possible below it. The second condition is important because the NII-Composites are an indicator of where the mixed population lies in the OB-I diagram. Therefore, if we want to create a separation line that includes solely galaxies whose main emission comes exclusively from SF, we need to exclude all components that might be mixed.

With these considerations in mind, the best fit we found was the following hyperbolic expression:
\begin{multline}
    \log\left( \frac{\oiii}{\hbeta} \right) = \frac{-1.0\pm0.2}{\log(\text{EW}(\hbeta)) - (0.68\pm0.09)} \\ + (1.53\pm0.12)
    \label{eq:obi population limit}
\end{multline}
The results of which can be seen in Fig. \ref{fig:model fit}.

At higher values of the EW of $\hbeta$, this second separation line is extremely similar to the empirical line we calculated (Eq. \ref{eq:hbeta limit}), further providing evidence that in this regime the separation between SF and AGN galaxies is clear.

\begin{figure}[h]
    \centering
    \includegraphics[width=1.0\linewidth]{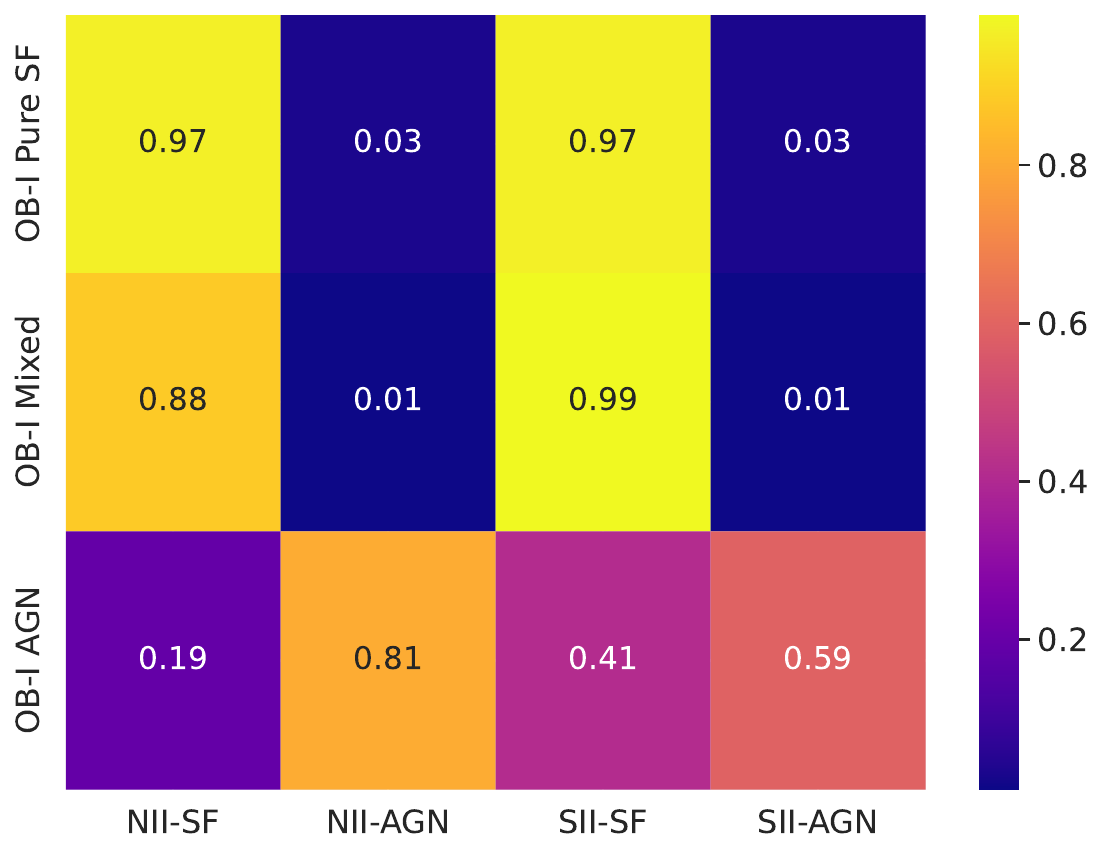}
    \caption{Match between the OB-I classes and the NII and SII diagram regions. This tells us how many galaxies in the OB-I classification belongs to the NII or SII classes, e.g., if we look at the top left square, we can see that $97\%$ of OB-I Pure SF are in the SF region of the NII diagram.}
    \label{fig:model cmatrix}
\end{figure}

In order to test if the pure SF and mixed population are what we claim them to be, we can look once again at the NII and SII diagrams. From the top panels of Fig. \ref{fig:obi class bpt} we can see that the Pure SF population of the OB-I diagram fits quite nicely in the limits of the NII diagram SF region, with only $3\%$ of the OB-I Pure SF galaxies crossing over into the Composite and AGN territories. This implies our modelled function of separation based on the \cite{stasinska01} models gives us a good estimate of galaxies which are only forming stars and have no AGN influence - taking into account a caveat that the low-metallicity regime is more complicated than both these diagrams make it seem. Looking at the bottom panels of Fig \ref{fig:obi class bpt}, we can see that the same is true for the SII diagram, where again only $3\%$ of the OB-I Pure SF population crosses over into the AGN region. This reinforces the point that this population is made up of galaxies whose primary source of excitation comes from star-formation, and essentially nothing else.

The mixed population is in line with what we expect it to be: a mix of galaxies that possess a combination of emissions from star-formation and AGN, though the SF component still seems to dominate over the AGN one. Evidence for this comes from their position in the NII and SII diagrams, where most of them are placed in the SF region (with $88\%$ in the NII-SF region and nearly $100\%$ in the SII-SF region). In Fig. \ref{fig:model cmatrix}, we can look at these values with more detail but, overall, with the addition of Eq. \ref{eq:obi population limit}, we can clearly see that we are able to improve the classification of galaxies that are purely SF, especially when compared to the NII diagram, with the detriment that the SII diagram seems to suffer from some loss in the classification. The AGN population does not suffer any alteration.

However, implications of the location of these two populations show that the OB-I diagram (including both Eqs. \ref{eq:hbeta limit} and \ref{eq:obi population limit}) is still imperfect at separating the mixed galaxy type. Nevertheless, we can claim that it is a good diagnosis tool of AGNs in the Local Universe and, through analysis of theoretical models, can distinguish galaxies whose main source of emission comes exclusively from star-formation. Any galaxy that has some sort of mixed regime is ambiguous, as it becomes very difficult to tell what dominates their emission, the star-formation or the AGN activity.

This discussion is limited to the Local Universe. In the next section, we discuss how the OB-I diagram reacts to higher redshift galaxies, how they evolve in the diagram, and what we can extract from it.

\section{Higher redshift landscape}\label{section:highz}

It is known that line ratio diagrams suffer from a `cosmic shift', where galaxies are placed in different positions in the parameter space depending on the redshift. The cause for this is still unknown, with explanations ranging from higher ionisation parameters of the gas in galaxies at higher redshifts (e.g. \citealt{brinchmann_2008,kewley_2013}), different contributions of the ionised gas and abundance ratios (e.g. \citealt{shapley_2015,cowie_2016}), differences in metallicity allowing for stronger emission lines (e.g. \citealt{steidel_2014,steidel_2016}), to the effects of shocks \citep{brinchmann_23}, or stellar rotation and binarity \citep{Bian_2020}. 

This implies that extrapolating the OB-I diagram from the Local Universe to a higher redshift is uncertain, as the `cosmic shift' can deeply impact this diagram and, therefore, our empirical separation line. To measure this impact, we need to analyse the location of galaxies at higher redshifts in the OB-I diagram.

\begin{figure*}[h]
    \centering
    \includegraphics[width=1.0\linewidth]{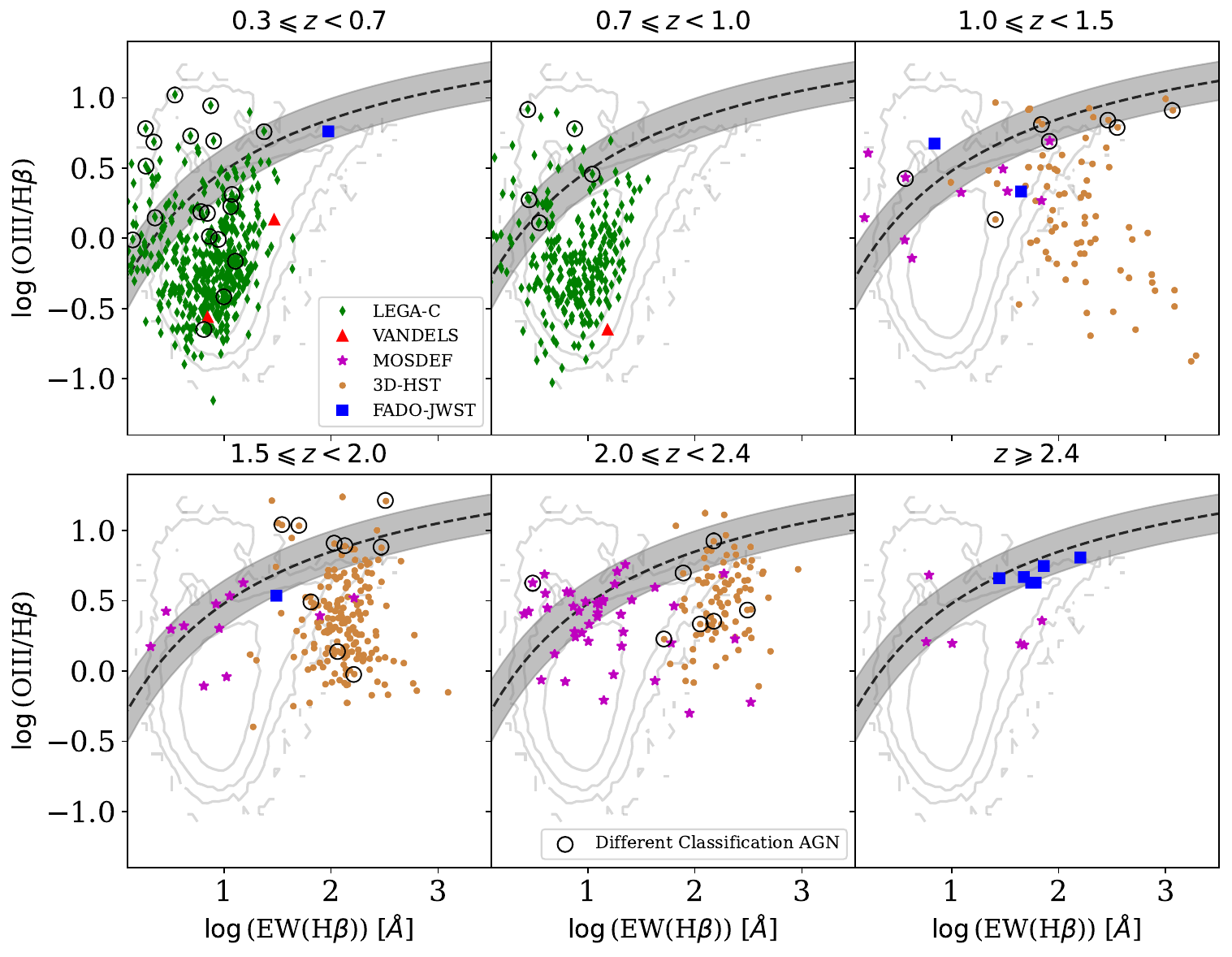}
    \caption{OB-I diagram of the higher redshift sample. The green diamonds represent the LEGA-C sample, the red triangles the VANDELS sample, the orange circles the 3D-HST sample, the magenta stars the MOSDEF sample and the blue squares the FADO-JWST sample. The encircled symbols represent galaxies that are classified as AGN by a separate classification scheme (see Appendix \ref{appendix:highz class}). The grey contours represent the FADO-SDSS sample. The black dashed line is the separation line we defined in Eq. \ref{eq:hbeta limit}. Above each plot we have the redshift bins defined.}
    \label{fig:highz obi}
\end{figure*}

Our high redshift sample was defined earlier in this work (see Sect. \ref{section:sample}), where we have $1\,152$ galaxies in the redshift range of $0.3 < z < 2.7$, and includes data from LEGA-C, VANDELS, 3D-HST,  MOSDEF and FADO-JWST. This sample is by no means exhaustive, but it will provide an idea of what the limitations of the OB-I diagram are at higher redshifts. 

Each of these datasets has an AGN selection system, which is what we used to compare with the OB-I diagram. As described in Sect. \ref{section:sample}, the galaxies we selected from the VANDELS survey had no AGN flag attached, so they are very likely SF. Furthermore, the  FADO-JWST galaxies were all selected to be SF galaxies.

When it comes to LEGA-C, 3D-HST and MOSDEF, the process of selecting AGNs is more complicated. We describe the respective selection for each of these datasets in detail in Appendix \ref{appendix:highz class}, but we present here a brief explanation. For LEGA-C, we made use of the \texttt{flag\_spec} column, which detects 24 AGNs. For 3D-HST, we made use of IRAC photometry and the \citet{donley2012} selection criteria, and found 20 galaxies to be AGN. For MOSDEF, we made use of the NII diagram and the \cite{kewley_2013} selection criteria, and found 3 AGNs. All remaining galaxies from these datasets are assumed to be SF.

In Fig. \ref{fig:highz obi}, we can see the whole higher redshift sample on the OB-I diagram, separated into six different redshift bins, with AGNs classified from other schemes highlighted with a black circle. In Table \ref{tab:obihighz}, we can see the information of these different classifications summed up, with how many AGNs match the OB-I classification, and how many AGNs in total the OB-I actually identifies. We can see that the OB-I diagram can track down approximately $50\%$ of AGNs found by different classification schemes, without the need for significant adjustments in our empirical line.

\begin{table}[h]
\centering
\caption{AGNs found by each higher redshift dataset considered in this work, by the OB-I diagram, and the match between them.}
\label{tab:obihighz}
\begin{tabular}{c|ccc}
Classification & AGNs Found & Match & OB-I AGNs Found \\ \hline
LEGA-C         & 24   & 13  & 87    \\
VANDELS        & 0    & 0   & 0     \\
MOSDEF         & 3    & 2   & 24     \\
3D-HST         & 20   & 7   & 24    \\
FADO-JWST      & 0    & 0   & 1     \\ \hline
Total          & 47   & 22  & 136    \\ \hline
\end{tabular}
\end{table}

It is of note that, as we move on from the lowest redshift to the highest redshift bins, the number of identified AGNs is lower and lower, in both the OB-I diagram and different galaxy classification schemes. This is likely due to the scarcity of sources observed and classified at redshifts above $z > 2.4$.

\begin{figure*}
    \centering
    \includegraphics[width=1.0\linewidth]{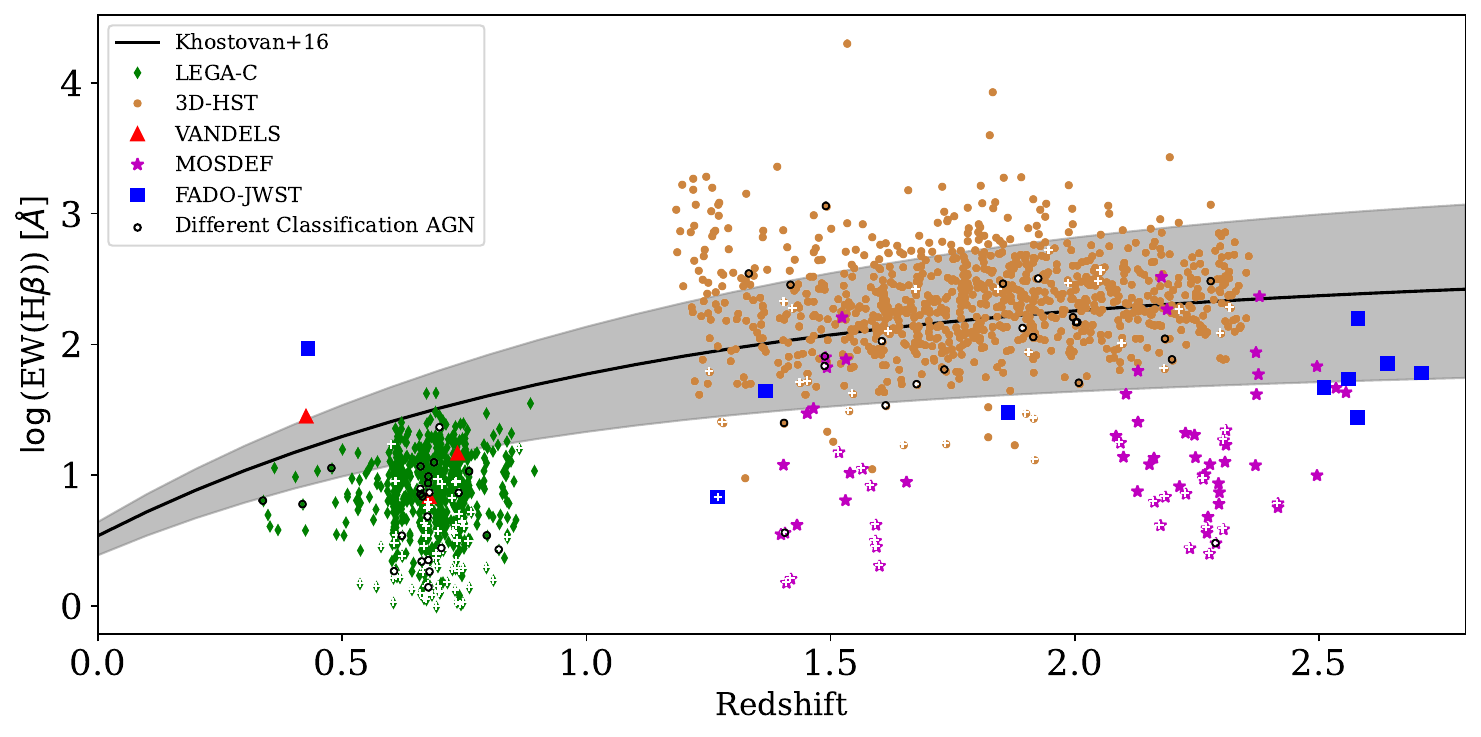}
    \caption{Evolution of the EW($\hbeta$) with redshift. The black line and gray shaded area are the expected evolution from \citet{khostovan2016}. The white crosses represent the OB-I AGNs. All remaining elements are the same as in Fig. \ref{fig:highz obi}.}
    \label{fig:ewhb vs z}
\end{figure*}

In Fig. \ref{fig:ewhb vs z}, we compare the galaxies in our higher-redshift sample with the expected redshift evolution of EW($\hbeta$) from \cite{khostovan2016}. There are two caveats in the aforementioned work that are worth discussing: first, they used galaxies that were both SF and AGN and secondly, these were based on EW measured using photometric filters, which even with narrowband could include effects from the $\oiii$ emission line. Regarding the first point, this does not cause any issues for us, since our sample is also a mixture of SF galaxies and AGNs.
Regarding the second point, this could be a problem, but the authors used data from several datasets to constrain the fit, allowing it to be compared with other methods. Another point to add is that, as mentioned before, FADO measures EWs differently when compared to other methods, but even taking into account this effect, the difference is about 0.1 dex, not enough to change the location of these galaxies significantly.

Overall, 53\% of galaxies are included in the expected redshift evolution of EW($\hbeta$), showing that more than half of our sample still follows the trend of increasing EW with redshift. It is of note that AGNs, from the OB-I or the aforementioned classification schemes (in white crosses and black circles, respectively), show only a slight trend with redshift, and remain at low EW values. This is something we expect, as we explained in the end of Sect. \ref{section:results}.

The flux ratio of $\oiii/\hbeta$ appears to increase as we increase the redshift as well, even in the limited sample we possess. This is in line with recent galaxy samples obtained using JWST, such as the Assembly of Ultradeep Rest-optical Observations Revealing Astrophysics (AURORA) survey, where a subtle increase in the flux ratio is expected between $1.4 < z < 4.0$ \citep{shapley25}, where most of our galaxies in the higher-redshift sample lie.

Another important facet to mention is that we are assuming that the classification from the datasets in our higher-redshift sample are correctly classifying their galaxies. All of the classification methods mentioned in Appendix \ref{appendix:highz class} separate galaxies in two types (SF and AGN), but the truth is that many of these likely lie in the middle, akin to the NII-Composites in the Local Universe. Our current paradigm in the Cosmic Noon lacks the density of detected sources like in the Local Universe, so differentiating a third population at higher redshifts can only done with, for example, spatially resolved and high resolution data (e.g. \citealt{lam25}). With this in mind, we still argue that, despite all of these uncertainties, the OB-I diagram would add an element to dispel the differences between galaxies that have mixed SF and AGN influence, from the purely AGNs, since galaxies do not seem to evolve significantly with redshift in the range we are studying.

With all these considerations in mind, we argue that the OB-I diagram, at least our empirical separation of galaxies, appears to be resistant to the `cosmic shift' that plagues most optical classification schemes. It has some success at classifying AGNs that other classification schemes identify, though there are still some that fall below our empirical separation line. Even with this setback (which we can clearly see in the $1 < z < 1.5$ bin), the fact that these AGNs classified by other methods still lie in the uncertainty region of our empirical line is very encouraging. Furthermore, the overwhelming majority of the galaxies are correctly classified as SF-dominated galaxies across all redshifts, and there is no need for major adjustments, unlike, for example the SF-AGN line in \cite{kewley_2013}. This can make the OB-I diagram and the empirical separation line we created a valuable tool, especially at redshifts between $1.7 < z < 2.7$, where the $\halpha$ and $\nii$ lines disappear out of the range of most optical spectrographs, leaving us with very few emission lines to work with. However, we need larger samples to further test these conclusions.

\section{Conclusions and Future Work}\label{section:conclusion}

Over the course of this work we explored the OB-I diagram, which compares the EW of $\hbeta$ and the emission line ratio of $\oiii/\hbeta$, in order to understand its properties and evolution. We selected $162\,239$ galaxies from the SDSS, LEGA-C, VANDELS, 3D-HST, MOSDEF and JADES surveys in order to understand the properties and evolution of the OB-I diagram (see Table \ref{tab:selection} and Fig. \ref{fig:all_redshift}).

The FADO-SDSS sample was first plotted on the OB-I diagram, as an analogue population for the Local Universe, and two clusters were apparent in the classification scheme: one dispersed in a `cloud' and another concentrated in an apparent `teardrop' shape. To separate these two regions, we used the NII diagram (comparing $[\text{NII}]\lambda6584/\text{H}\alpha$ with $[\text{OIII}]\lambda5007/\text{H}\beta$) to create an empirical line, given by Eq. \ref{eq:hbeta limit}, that essentially distinguishes between galaxies with a strong AGN contribution and galaxies with SF-dominated emission (see Fig. \ref{fig:obi_separation}). We further compared this classification with the SII diagram, and found that $99\%$ of the OB-I SF galaxies are below the \cite{kewley_2001} theoretical `maximum starburst' line and that $60\%$ of the OB-I AGNs were above this same line (see Fig. \ref{fig:bpt obi}). This tells us that there is a remarkably good agreement with the SII diagram classification, meaning that our empirical classification scheme is in line with the current paradigm of galaxy classification.

We further compared the OB-I diagram with the stellar evolution models of \cite{stasinska01}, which can be seen in Fig. \ref{fig:models}. From these models we reached two conclusions: 1) at higher values of EW of $\hbeta$ (above $\approx 10^{2} \: \AA$), the models agree with our empirical classification line, meaning that the theory agrees with our separation between SF galaxies and AGNs, and 2) below this value, the models show a sharp drop in the $\oiii/\hbeta$ emission line ratio, implying the presence of a third population. This leads to an additional separation for separating the SF-dominated population in the OB-I diagram: one is made up of galaxies with only SF emission and another composed of galaxies with SF and AGN emission mixed together. We created a line, given by Eq. \ref{eq:obi population limit}, to separate these two populations, although it is of note that there are many galaxies in the mixed population that are likely SF only, with no AGN contribution -- however, we argue that galaxies below this line are mostly purely SF (see Fig. \ref{fig:obi class bpt}).

Finally, and most importantly, we compared the OB-I diagram with the higher redshift sample ($0.3 < z < 2.7$), which includes the LEGA-C, VANDELS, MOSDEF, 3D-HST and JADES surveys (Fig. \ref{fig:highz obi}). We found that our empirical classification scheme appears to be resistant to the `cosmic shift' that plagues most classification schemes in the optical \citep{brinchmann_2008, steidel_2014, steidel_2016, shapley_2015, cowie_2016, brinchmann_23,Bian_2020}. From flags and other classification schemes, we find that, at higher redshifts, the OB-I diagram correctly classifies half of the AGNs, with possibly no dependance on their redshift. With this in mind, we argue that the OB-I can be a valuable tool for classifying galaxies, with no need for modifications in separation lines according to redshift. In higher redshift sources where emission lines (such as $\halpha$ and $\nii$), required for other diagrams, are unavailable or hard to detect, the OB-I diagram acts as a first filter to understand the type of galaxies we are working with.

In the future, we plan on further exploring the redshift evolution of the OB-I diagram with more detail, by using data from future surveys, such as the MOONS Redshift-Intensive Survey Experiment (MOONRISE, \citealt{moons1,moons2,moon3}), the Prime Focus Spectrograph (PFS, \citealt{pfs}) and the WHT Enhanced Area Velocity Explorer (WEAVE, \citealt{weave}), as well as by generating theoretical models we can compare this data to.

\begin{acknowledgements}
This work was supported by Fundação para a Ciência e a Tecnologia (FCT) through the research grants PTDC/FIS-AST/29245/2017, UIDB/04434/2020 and UIDP/04434/2020. D.M.S. would like to acknowledge the meaningful discussions with Bruno Arsioli, who provided interesting insights to the paper. He would also like to acknowledge the patience and commitment of the team at the OAL and the IA, as well as the incredibly fruitful exchanges with C.L.. C.P. acknowledges support from DL 57/2016 (P2460) from the ‘Departamento de Física, Faculdade de Ciências da Universidade de Lisboa’. H.M. acknowledges support from the Fundação para a Ciência e a Tecnologia (FCT) through the PhD Fellowship 2022.12891.BD. R.C. acknowledges support from the Fundação para a Ciência e a Tecnologia (FCT) through the Fellowship PD/BD/150455/2019 (PhD:SPACE Doctoral Network PD/00040/2012) and POCH/FSE (EC). P.L. gratefully acknowledges support by the GEMINI ANID project No. 32240002.

Funding for the SDSS and SDSS-II has been provided by the Alfred P. Sloan Foundation, the Participating Institutions, the National Science Foundation, the U.S. Department of Energy, the National Aeronautics and Space Administration, the Japanese Monbukagakusho, the Max Planck Society, and the Higher Education Funding Council for England. The SDSS Web Site is http://www.sdss.org/. The SDSS is managed by the Astrophysical Research Consortium for the Participating Institutions. The Participating Institutions are the American Museum of Natural History, Astrophysical Institute Potsdam, University of Basel, University of Cambridge, Case Western Reserve University, University of Chicago, Drexel University, Fermilab, the Institute for Advanced Study, the Japan Participation Group, Johns Hopkins University, the Joint Institute for Nuclear Astrophysics, the Kavli Institute for Particle Astrophysics and Cosmology, the Korean Scientist Group, the Chinese Academy of Sciences (LAMOST), Los Alamos National Laboratory, the Max-Planck-Institute for Astronomy (MPIA), the Max-Planck-Institute for Astrophysics (MPA), New Mexico State University, Ohio State University, University of Pittsburgh, University of Portsmouth, Princeton University, the United States Naval Observatory, and the University of Washington. This work would also like to acknowledge the support from Matplotlib \citep{matplotlib}.

This work is based on observations taken by the 3D-HST Treasury Program (GO 12177 and 12328) with the NASA/ESA HST, which is operated by the Association of Universities for Research in Astronomy, Inc., under NASA contract NAS5-26555.

The UKIDSS project is defined in Lawrence et al (2007). Further details on the UDS can be found in \cite{almaini}. UKIDSS uses the UKIRT Wide Field Camera (WFCAM; \citealt{casali}). The photometric system is described in \cite{hewett06}, and the calibration is described in \cite{hodgkin09}. The pipeline processing and science archive are described in \cite{irwin04} and \cite{hambly08}.

This study makes use of data from AEGIS, a multiwavelength sky survey conducted with the Chandra, GALEX, Hubble, Keck, CFHT, MMT, Subaru, Palomar, Spitzer, VLA, and other telescopes and supported in part by the NSF, NASA, and the STFC.

This work is based, in part, on observations made with the Spitzer Space Telescope, which was operated by the Jet Propulsion Laboratory, California Institute of Technology under a contract with NASA.
\end{acknowledgements}

\bibliographystyle{aa}
\bibliography{bibliography.bib}

\begin{thebibliography}{109}
\expandafter\ifx\csname natexlab\endcsname\relax\def\natexlab#1{#1}\fi

\bibitem[{{Abazajian} {et~al.}(2009){Abazajian}, {Adelman-McCarthy}, {Ag{\"u}eros}, {Allam}, {Allende Prieto}, {An}, {Anderson}, {Anderson}, {Annis}, {Bahcall}, {Bailer-Jones}, {Barentine}, {Bassett}, {Becker}, {Beers}, {Bell}, {Belokurov}, {Berlind}, {Berman}, {Bernardi}, {Bickerton}, {Bizyaev}, {Blakeslee}, {Blanton}, {Bochanski}, {Boroski}, {Brewington}, {Brinchmann}, {Brinkmann}, {Brunner}, {Budav{\'a}ri}, {Carey}, {Carliles}, {Carr}, {Castander}, {Cinabro}, {Connolly}, {Csabai}, {Cunha}, {Czarapata}, {Davenport}, {de Haas}, {Dilday}, {Doi}, {Eisenstein}, {Evans}, {Evans}, {Fan}, {Friedman}, {Frieman}, {Fukugita}, {G{\"a}nsicke}, {Gates}, {Gillespie}, {Gilmore}, {Gonzalez}, {Gonzalez}, {Grebel}, {Gunn}, {Gy{\"o}ry}, {Hall}, {Harding}, {Harris}, {Harvanek}, {Hawley}, {Hayes}, {Heckman}, {Hendry}, {Hennessy}, {Hindsley}, {Hoblitt}, {Hogan}, {Hogg}, {Holtzman}, {Hyde}, {Ichikawa}, {Ichikawa}, {Im}, {Ivezi{\'c}}, {Jester}, {Jiang}, {Johnson}, {Jorgensen}, {Juri{\'c}}, {Kent}, {Kessler}, {Kleinman}, {Knapp},
  {Konishi}, {Kron}, {Krzesinski}, {Kuropatkin}, {Lampeitl}, {Lebedeva}, {Lee}, {Lee}, {French Leger}, {L{\'e}pine}, {Li}, {Lima}, {Lin}, {Long}, {Loomis}, {Loveday}, {Lupton}, {Magnier}, {Malanushenko}, {Malanushenko}, {Mandelbaum}, {Margon}, {Marriner}, {Mart{\'\i}nez-Delgado}, {Matsubara}, {McGehee}, {McKay}, {Meiksin}, {Morrison}, {Mullally}, {Munn}, {Murphy}, {Nash}, {Nebot}, {Neilsen}, {Newberg}, {Newman}, {Nichol}, {Nicinski}, {Nieto-Santisteban}, {Nitta}, {Okamura}, {Oravetz}, {Ostriker}, {Owen}, {Padmanabhan}, {Pan}, {Park}, {Pauls}, {Peoples}, {Percival}, {Pier}, {Pope}, {Pourbaix}, {Price}, {Purger}, {Quinn}, {Raddick}, {Re Fiorentin}, {Richards}, {Richmond}, {Riess}, {Rix}, {Rockosi}, {Sako}, {Schlegel}, {Schneider}, {Scholz}, {Schreiber}, {Schwope}, {Seljak}, {Sesar}, {Sheldon}, {Shimasaku}, {Sibley}, {Simmons}, {Sivarani}, {Allyn Smith}, {Smith}, {Smol{\v{c}}i{\'c}}, {Snedden}, {Stebbins}, {Steinmetz}, {Stoughton}, {Strauss}, {SubbaRao}, {Suto}, {Szalay}, {Szapudi}, {Szkody}, {Tanaka},
  {Tegmark}, {Teodoro}, {Thakar}, {Tremonti}, {Tucker}, {Uomoto}, {Vanden Berk}, {Vandenberg}, {Vidrih}, {Vogeley}, {Voges}, {Vogt}, {Wadadekar}, {Watters}, {Weinberg}, {West}, {White}, {Wilhite}, {Wonders}, {Yanny}, {Yocum}, {York}, {Zehavi}, {Zibetti}, \& {Zucker}}]{sdss}
{Abazajian}, K.~N., {Adelman-McCarthy}, J.~K., {Ag{\"u}eros}, M.~A., {et~al.} 2009, \apjs, 182, 543

\bibitem[{{Allen} {et~al.}(2008){Allen}, {Groves}, {Dopita}, {Sutherland}, \& {Kewley}}]{allen08}
{Allen}, M.~G., {Groves}, B.~A., {Dopita}, M.~A., {Sutherland}, R.~S., \& {Kewley}, L.~J. 2008, \apjs, 178, 20

\bibitem[{{Almaini} {et~al.}(2007){Almaini}, {Foucaud}, {Lane}, {Conselice}, {McLure}, {Cirasuolo}, {Dunlop}, {Smail}, \& {Simpson}}]{almaini}
{Almaini}, O., {Foucaud}, S., {Lane}, K., {et~al.} 2007, in Astronomical Society of the Pacific Conference Series, Vol. 379, Cosmic Frontiers, ed. N.~{Metcalfe} \& T.~{Shanks}, 163

\bibitem[{{Alongi} {et~al.}(1993){Alongi}, {Bertelli}, {Bressan}, {Chiosi}, {Fagotto}, {Greggio}, \& {Nasi}}]{Alongi_1993}
{Alongi}, M., {Bertelli}, G., {Bressan}, A., {et~al.} 1993, \aaps, 97, 851

\bibitem[{{Backhaus} {et~al.}(2022){Backhaus}, {Trump}, {Cleri}, {Simons}, {Momcheva}, {Papovich}, {Estrada-Carpenter}, {Finkelstein}, {Matharu}, {Ji}, {Weiner}, {Giavalisco}, \& {Jung}}]{ohno}
{Backhaus}, B.~E., {Trump}, J.~R., {Cleri}, N.~J., {et~al.} 2022, \apj, 926, 161

\bibitem[{{Baldwin} {et~al.}(1981){Baldwin}, {Phillips}, \& {Terlevich}}]{bpt}
{Baldwin}, J.~A., {Phillips}, M.~M., \& {Terlevich}, R. 1981, \pasp, 93, 5

\bibitem[{Bian {et~al.}(2020)Bian, Kewley, Groves, \& Dopita}]{Bian_2020}
Bian, F., Kewley, L.~J., Groves, B., \& Dopita, M.~A. 2020, Monthly Notices of the Royal Astronomical Society, 493, 580

\bibitem[{{Brammer} {et~al.}(2012){Brammer}, {van Dokkum}, {Franx}, {Fumagalli}, {Patel}, {Rix}, {Skelton}, {Kriek}, {Nelson}, {Schmidt}, {Bezanson}, {da Cunha}, {Erb}, {Fan}, {F{\"o}rster Schreiber}, {Illingworth}, {Labb{\'e}}, {Leja}, {Lundgren}, {Magee}, {Marchesini}, {McCarthy}, {Momcheva}, {Muzzin}, {Quadri}, {Steidel}, {Tal}, {Wake}, {Whitaker}, \& {Williams}}]{brammer_12}
{Brammer}, G.~B., {van Dokkum}, P.~G., {Franx}, M., {et~al.} 2012, \apjs, 200, 13

\bibitem[{{Bressan} {et~al.}(1993){Bressan}, {Fagotto}, {Bertelli}, \& {Chiosi}}]{Bressan_1993}
{Bressan}, A., {Fagotto}, F., {Bertelli}, G., \& {Chiosi}, C. 1993, \aaps, 100, 647

\bibitem[{{Brinchmann}(2023)}]{brinchmann_23}
{Brinchmann}, J. 2023, \mnras, 525, 2087

\bibitem[{{Brinchmann} {et~al.}(2004){Brinchmann}, {Charlot}, {White}, {Tremonti}, {Kauffmann}, {Heckman}, \& {Brinkmann}}]{brinchmann}
{Brinchmann}, J., {Charlot}, S., {White}, S.~D.~M., {et~al.} 2004, \mnras, 351, 1151

\bibitem[{{Brinchmann} {et~al.}(2008){Brinchmann}, {Pettini}, \& {Charlot}}]{brinchmann_2008}
{Brinchmann}, J., {Pettini}, M., \& {Charlot}, S. 2008, \mnras, 385, 769

\bibitem[{{Bruzual} \& {Charlot}(2003)}]{Bruzual_2003}
{Bruzual}, G. \& {Charlot}, S. 2003, \mnras, 344, 1000

\bibitem[{{Bunker} {et~al.}(2024){Bunker}, {Cameron}, {Curtis-Lake}, {Jakobsen}, {Carniani}, {Curti}, {Witstok}, {Maiolino}, {D'Eugenio}, {Looser}, {Willott}, {Bonaventura}, {Hainline}, {{\"U}bler}, {Willmer}, {Saxena}, {Smit}, {Alberts}, {Arribas}, {Baker}, {Baum}, {Bhatawdekar}, {Bowler}, {Boyett}, {Charlot}, {Chen}, {Chevallard}, {Circosta}, {DeCoursey}, {de Graaff}, {Egami}, {Eisenstein}, {Endsley}, {Ferruit}, {Giardino}, {Hausen}, {Helton}, {Hviding}, {Ji}, {Johnson}, {Jones}, {Kumari}, {Laseter}, {L{\"u}tzgendorf}, {Maseda}, {Nelson}, {Parlanti}, {Perna}, {Rauscher}, {Rawle}, {Rix}, {Rieke}, {Robertson}, {Rodr{\'\i}guez Del Pino}, {Sandles}, {Scholtz}, {Sharpe}, {Skarbinski}, {Stark}, {Sun}, {Tacchella}, {Topping}, {Villanueva}, {Wallace}, {Williams}, \& {Woodrum}}]{jades1}
{Bunker}, A.~J., {Cameron}, A.~J., {Curtis-Lake}, E., {et~al.} 2024, \aap, 690, A288

\bibitem[{{Byler} {et~al.}(2017){Byler}, {Dalcanton}, {Conroy}, \& {Johnson}}]{byler17}
{Byler}, N., {Dalcanton}, J.~J., {Conroy}, C., \& {Johnson}, B.~D. 2017, \apj, 840, 44

\bibitem[{Calzetti {et~al.}(2000)Calzetti, Armus, Bohlin, Kinney, Koornneef, \& Storchi-Bergmann}]{calzetti_2000}
Calzetti, D., Armus, L., Bohlin, R.~C., {et~al.} 2000, The Astrophysical Journal, 533, 682

\bibitem[{{Cardelli} {et~al.}(1989){Cardelli}, {Clayton}, \& {Mathis}}]{Cardelli_1989}
{Cardelli}, J.~A., {Clayton}, G.~C., \& {Mathis}, J.~S. 1989, \apj, 345, 245

\bibitem[{{Cardoso} {et~al.}(2022){Cardoso}, {Gomes}, {Papaderos}, {Pappalardo}, {Miranda}, {Paulino-Afonso}, {Afonso}, \& {Lagos}}]{cardoso}
{Cardoso}, L. S.~M., {Gomes}, J.~M., {Papaderos}, P., {et~al.} 2022, \aap, 667, A11

\bibitem[{{Casado} {et~al.}(2015){Casado}, {Ascasibar}, {Gavil{\'a}n}, {Terlevich}, {Terlevich}, {Hoyos}, \& {D{\'\i}az}}]{casado15}
{Casado}, J., {Ascasibar}, Y., {Gavil{\'a}n}, M., {et~al.} 2015, \mnras, 451, 888

\bibitem[{{Casali} {et~al.}(2007){Casali}, {Adamson}, {Alves de Oliveira}, {Almaini}, {Burch}, {Chuter}, {Elliot}, {Folger}, {Foucaud}, {Hambly}, {Hastie}, {Henry}, {Hirst}, {Irwin}, {Ives}, {Lawrence}, {Laidlaw}, {Lee}, {Lewis}, {Lunney}, {McLay}, {Montgomery}, {Pickup}, {Read}, {Rees}, {Robson}, {Sekiguchi}, {Vick}, {Warren}, \& {Woodward}}]{casali}
{Casali}, M., {Adamson}, A., {Alves de Oliveira}, C., {et~al.} 2007, \aap, 467, 777

\bibitem[{{Chabrier}(2003)}]{Chabrier_2003}
{Chabrier}, G. 2003, \apjl, 586, L133

\bibitem[{{Cid Fernandes} {et~al.}(2011){Cid Fernandes}, {Stasi{\'n}ska}, {Mateus}, \& {Vale Asari}}]{cidfernandes}
{Cid Fernandes}, R., {Stasi{\'n}ska}, G., {Mateus}, A., \& {Vale Asari}, N. 2011, \mnras, 413, 1687

\bibitem[{{Cid Fernandes} {et~al.}(2010){Cid Fernandes}, {Stasi{\'n}ska}, {Schlickmann}, {Mateus}, {Vale Asari}, {Schoenell}, \& {Sodr{\'e}}}]{cidfernandes_2010}
{Cid Fernandes}, R., {Stasi{\'n}ska}, G., {Schlickmann}, M.~S., {et~al.} 2010, \mnras, 403, 1036

\bibitem[{{Cirasuolo} {et~al.}(2020){Cirasuolo}, {Fairley}, {Rees}, {Gonzalez}, {Taylor}, {Maiolino}, {Afonso}, {Evans}, {Flores}, {Lilly}, {Oliva}, {Paltani}, {Vanzi}, {Abreu}, {Accardo}, {Adams}, {{\'A}lvarez M{\'e}ndez}, {Amans}, {Amarantidis}, {Atek}, {Atkinson}, {Banerji}, {Barrett}, {Barrientos}, {Bauer}, {Beard}, {B{\'e}chet}, {Belfiore}, {Bellazzini}, {Benoist}, {Best}, {Biazzo}, {Black}, {Boettger}, {Bonifacio}, {Bowler}, {Bragaglia}, {Brierley}, {Brinchmann}, {Brinkmann}, {Buat}, {Buitrago}, {Burgarella}, {Burningham}, {Buscher}, {Cabral}, {Caffau}, {Cardoso}, {Carnall}, {Carollo}, {Castillo}, {Castignani}, {Catelan}, {Cicone}, {Cimatti}, {Cioni}, {Clementini}, {Cochrane}, {Coelho}, {Colling}, {Contini}, {Contreras}, {Conzelmann}, {Cresci}, {Cropper}, {Cucciati}, {Cullen}, {Cumani}, {Curti}, {Da Silva}, {Daddi}, {Dalessandro}, {Dalessio}, {Dauvin}, {Davidson}, {de Laverny}, {Delplancke-Str{\"o}bele}, {De Lucia}, {Del Vecchio}, {Dessauges-Zavadsky}, {Di Matteo}, {Dole}, {Drass}, {Dunlop},
  {D{\"u}nner}, {Eales}, {Ellis}, {Enriques}, {Fasola}, {Ferguson}, {Ferruzzi}, {Fisher}, {Flores}, {Fontana}, {Forchi}, {Francois}, {Franzetti}, {Gargiulo}, {Garilli}, {Gaudemard}, {Gieles}, {Gilmore}, {Ginolfi}, {Gomes}, {Guinouard}, {Gutierrez}, {Haigron}, {Hammer}, {Hammersley}, {Haniff}, {Harrison}, {Haywood}, {Hill}, {Hubin}, {Humphrey}, {Ibata}, {Infante}, {Ives}, {Ivison}, {Iwert}, {Jablonka}, {Jakob}, {Jarvis}, {King}, {Kneib}, {Laporte}, {Lawrence}, {Lee}, {Li Causi}, {Lorenzoni}, {Lucatello}, {Luco}, {Macleod}, {Magliocchetti}, {Magrini}, {Mainieri}, {Maire}, {Mannucci}, {Martin}, {Matute}, {Maurogordato}, {McGee}, {Mcleod}, {McLure}, {McMahon}, {Melse}, {Messias}, {Mucciarelli}, {Nisini}, {Nix}, {Norberg}, {Oesch}, {Oliveira}, {Origlia}, {Padilla}, {Palsa}, {Pancino}, {Papaderos}, {Pappalardo}, {Parry}, {Pasquini}, {Peacock}, {Pedichini}, {Pello}, {Peng}, {Pentericci}, {Pfuhl}, {Piazzesi}, {Popovic}, {Pozzetti}, {Puech}, {Puzia}, {Raichoor}, {Randich}, {Recio-Blanco}, {Reis}, {Reix}, {Renzini},
  {Rodrigues}, {Rojas}, {Rojas-Arriagada}, {Rota}, {Royer}, {Sacco}, {Sanchez-Janssen}, {Sanna}, {Santos}, {Sarzi}, {Schaerer}, {Schiavon}, {Schnell}, {Schultheis}, {Scodeggio}, {Serjeant}, {Shen}, {Simmonds}, {Smoker}, {Sobral}, {Sordet}, {Sp{\'e}rone}, {Strachan}, {Sun}, {Swinbank}, {Tait}, {Tereno}, {Tojeiro}, {Torres}, {Tosi}, {Tozzi}, {Tresiter}, {Valenti}, {Valenzuela Navarro}, {Vanzella}, {Vergani}, {Verhamme}, {Vernet}, {Vignali}, {Vinther}, {Von Dran}, {Waring}, {Watson}, {Wild}, {Willesme}, {Woodward}, {Wuyts}, {Yang}, {Zamorani}, {Zoccali}, {Bluck}, \& {Trussler}}]{moons2}
{Cirasuolo}, M., {Fairley}, A., {Rees}, P., {et~al.} 2020, The Messenger, 180, 10

\bibitem[{{Cirasuolo} \& {MOONS Consortium}(2016)}]{moons1}
{Cirasuolo}, M. \& {MOONS Consortium}. 2016, in Astronomical Society of the Pacific Conference Series, Vol. 507, Multi-Object Spectroscopy in the Next Decade: Big Questions, Large Surveys, and Wide Fields, ed. I.~{Skillen}, M.~{Balcells}, \& S.~{Trager}, 109

\bibitem[{{Cowie} {et~al.}(2016){Cowie}, {Barger}, \& {Songaila}}]{cowie_2016}
{Cowie}, L.~L., {Barger}, A.~J., \& {Songaila}, A. 2016, \apj, 817, 57

\bibitem[{{Curti} {et~al.}(2017){Curti}, {Cresci}, {Mannucci}, {Marconi}, {Maiolino}, \& {Esposito}}]{curti17}
{Curti}, M., {Cresci}, G., {Mannucci}, F., {et~al.} 2017, \mnras, 465, 1384

\bibitem[{{Curti} {et~al.}(2020){Curti}, {Mannucci}, {Cresci}, \& {Maiolino}}]{curti20}
{Curti}, M., {Mannucci}, F., {Cresci}, G., \& {Maiolino}, R. 2020, \mnras, 491, 944

\bibitem[{{Dalton}(2016)}]{weave}
{Dalton}, G. 2016, in Astronomical Society of the Pacific Conference Series, Vol. 507, Multi-Object Spectroscopy in the Next Decade: Big Questions, Large Surveys, and Wide Fields, ed. I.~{Skillen}, M.~{Balcells}, \& S.~{Trager}, 97

\bibitem[{{Davis} {et~al.}(2007){Davis}, {Guhathakurta}, {Konidaris}, {Newman}, {Ashby}, {Biggs}, {Barmby}, {Bundy}, {Chapman}, {Coil}, {Conselice}, {Cooper}, {Croton}, {Eisenhardt}, {Ellis}, {Faber}, {Fang}, {Fazio}, {Georgakakis}, {Gerke}, {Goss}, {Gwyn}, {Harker}, {Hopkins}, {Huang}, {Ivison}, {Kassin}, {Kirby}, {Koekemoer}, {Koo}, {Laird}, {Le Floc'h}, {Lin}, {Lotz}, {Marshall}, {Martin}, {Metevier}, {Moustakas}, {Nandra}, {Noeske}, {Papovich}, {Phillips}, {Rich}, {Rieke}, {Rigopoulou}, {Salim}, {Schiminovich}, {Simard}, {Smail}, {Small}, {Weiner}, {Willmer}, {Willner}, {Wilson}, {Wright}, \& {Yan}}]{aegis}
{Davis}, M., {Guhathakurta}, P., {Konidaris}, N.~P., {et~al.} 2007, \apjl, 660, L1

\bibitem[{{Denicol{\'o}} {et~al.}(2002){Denicol{\'o}}, {Terlevich}, \& {Terlevich}}]{denicolo_2002}
{Denicol{\'o}}, G., {Terlevich}, R., \& {Terlevich}, E. 2002, \mnras, 330, 69

\bibitem[{{DESI Collaboration} {et~al.}(2024){DESI Collaboration}, {Adame}, {Aguilar}, {Ahlen}, {Alam}, {Aldering}, {Alexander}, {Alfarsy}, {Allende Prieto}, {Alvarez}, {Alves}, {Anand}, {Andrade-Oliveira}, {Armengaud}, {Asorey}, {Avila}, {Aviles}, {Bailey}, {Balaguera-Antol{\'\i}nez}, {Ballester}, {Baltay}, {Bault}, {Bautista}, {Behera}, {Beltran}, {BenZvi}, {Beraldo e Silva}, {Bermejo-Climent}, {Berti}, {Besuner}, {Beutler}, {Bianchi}, {Blake}, {Blum}, {Bolton}, {Brieden}, {Brodzeller}, {Brooks}, {Brown}, {Buckley-Geer}, {Burtin}, {Cabayol-Garcia}, {Cai}, {Canning}, {Cardiel-Sas}, {Carnero Rosell}, {Castander}, {Cervantes-Cota}, {Chabanier}, {Chaussidon}, {Chaves-Montero}, {Chen}, {Chen}, {Chuang}, {Claybaugh}, {Cole}, {Cooper}, {Cuceu}, {Davis}, {Dawson}, {de Belsunce}, {de la Cruz}, {de la Macorra}, {Della Costa}, {de Mattia}, {Demina}, {Demirbozan}, {DeRose}, {Dey}, {Dey}, {Dhungana}, {Ding}, {Ding}, {Doel}, {Doshi}, {Douglass}, {Edge}, {Eftekharzadeh}, {Eisenstein}, {Elliott}, {Ereza}, {Escoffier},
  {Fagrelius}, {Fan}, {Fanning}, {Fawcett}, {Ferraro}, {Flaugher}, {Font-Ribera}, {Forero-Romero}, {Forero-S{\'a}nchez}, {Frenk}, {G{\"a}nsicke}, {Garc{\'\i}a}, {Garc{\'\i}a-Bellido}, {Garcia-Quintero}, {Garrison}, {Gil-Mar{\'\i}n}, {Golden-Marx}, {Gontcho A Gontcho}, {Gonzalez-Morales}, {Gonzalez-Perez}, {Gordon}, {Graur}, {Green}, {Gruen}, {Guy}, {Hadzhiyska}, {Hahn}, {Han}, {Hanif}, {Herrera-Alcantar}, {Honscheid}, {Hou}, {Howlett}, {Huterer}, {Ir{\v{s}}i{\v{c}}}, {Ishak}, {Jacques}, {Jana}, {Jiang}, {Jimenez}, {Jing}, {Joudaki}, {Joyce}, {Jullo}, {Juneau}, {Kara{\c{c}}ayl{\i}}, {Karim}, {Kehoe}, {Kent}, {Khederlarian}, {Kim}, {Kirkby}, {Kisner}, {Kitaura}, {Kizhuprakkat}, {Kneib}, {Koposov}, {Kov{\'a}cs}, {Kremin}, {Krolewski}, {L'Huillier}, {Lahav}, {Lambert}, {Lamman}, {Lan}, {Landriau}, {Lang}, {Lange}, {Lasker}, {Leauthaud}, {Le Guillou}, {Levi}, {Li}, {Linder}, {Lyons}, {Magneville}, {Manera}, {Manser}, {Margala}, {Martini}, {McDonald}, {Medina}, {Medina-Varela}, {Meisner}, {Mena-Fern{\'a}ndez},
  {Meneses-Rizo}, {Mezcua}, {Miquel}, {Montero-Camacho}, {Moon}, {Moore}, {Moustakas}, {Mueller}, {Mundet}, {Mu{\~n}oz-Guti{\'e}rrez}, {Myers}, {Nadathur}, {Napolitano}, {Neveux}, {Newman}, {Nie}, {Nikutta}, {Niz}, {Norberg}, {Noriega}, {Paillas}, {Palanque-Delabrouille}, {Palmese}, {Pan}, {Parkinson}, {Penmetsa}, {Percival}, {P{\'e}rez-Fern{\'a}ndez}, {P{\'e}rez-R{\`a}fols}, {Pieri}, {Poppett}, {Porredon}, {Pothier}, {Prada}, {Pucha}, {Raichoor}, {Ram{\'\i}rez-P{\'e}rez}, {Ramirez-Solano}, {Rashkovetskyi}, {Ravoux}, {Rocher}, {Rockosi}, {Ross}, {Rossi}, {Ruggeri}, {Ruhlmann-Kleider}, {Sabiu}, {Said}, {Saintonge}, {Samushia}, {Sanchez}, {Saulder}, {Schaan}, {Schlafly}, {Schlegel}, {Scholte}, {Schubnell}, {Seo}, {Shafieloo}, {Sharples}, {Sheu}, {Silber}, {Sinigaglia}, {Siudek}, {Slepian}, {Smith}, {Soumagnac}, {Sprayberry}, {Stephey}, {Su{\'a}rez-P{\'e}rez}, {Sun}, {Tan}, {Tarl{\'e}}, {Tojeiro}, {Ure{\~n}a-L{\'o}pez}, {Vaisakh}, {Valcin}, {Valdes}, {Valluri}, {Vargas-Maga{\~n}a}, {Variu}, {Verde}, {Walther},
  {Wang}, {Wang}, {Weaver}, {Weaverdyck}, {Wechsler}, {White}, {Xie}, {Yang}, {Y{\`e}che}, {Yu}, {Yuan}, {Zhang}, {Zhang}, {Zhao}, {Zheng}, {Zhou}, {Zhou}, {Zou}, {Zou}, \& {Zu}}]{desi_edr}
{DESI Collaboration}, {Adame}, A.~G., {Aguilar}, J., {et~al.} 2024, \aj, 168, 58

\bibitem[{{Dickinson} {et~al.}(2003){Dickinson}, {Giavalisco}, \& {GOODS Team}}]{goods-s}
{Dickinson}, M., {Giavalisco}, M., \& {GOODS Team}. 2003, in The Mass of Galaxies at Low and High Redshift, ed. R.~{Bender} \& A.~{Renzini}, 324

\bibitem[{{Donley} {et~al.}(2012){Donley}, {Koekemoer}, {Brusa}, {Capak}, {Cardamone}, {Civano}, {Ilbert}, {Impey}, {Kartaltepe}, {Miyaji}, {Salvato}, {Sanders}, {Trump}, \& {Zamorani}}]{donley2012}
{Donley}, J.~L., {Koekemoer}, A.~M., {Brusa}, M., {et~al.} 2012, \apj, 748, 142

\bibitem[{{Eisenstein} {et~al.}(2023{\natexlab{a}}){Eisenstein}, {Johnson}, {Robertson}, {Tacchella}, {Hainline}, {Jakobsen}, {Maiolino}, {Bonaventura}, {Bunker}, {Cameron}, {Cargile}, {Curtis-Lake}, {Hausen}, {Pusk{\'a}s}, {Rieke}, {Sun}, {Willmer}, {Willott}, {Alberts}, {Arribas}, {Baker}, {Baum}, {Bhatawdekar}, {Carniani}, {Charlot}, {Chen}, {Chevallard}, {Curti}, {DeCoursey}, {D'Eugenio}, {de Graaff}, {Egami}, {Helton}, {Ji}, {Jones}, {Kumari}, {L{\"u}tzgendorf}, {Laseter}, {Looser}, {Lyu}, {Maseda}, {Nelson}, {Parlanti}, {Rauscher}, {Rawle}, {Rieke}, {Rix}, {Rujopakarn}, {Sandles}, {Saxena}, {Scholtz}, {Sharpe}, {Shivaei}, {Simmonds}, {Smit}, {Topping}, {{\"U}bler}, {Venturi}, {Williams}, {Witstok}, \& {Woodrum}}]{jades5}
{Eisenstein}, D.~J., {Johnson}, B.~D., {Robertson}, B., {et~al.} 2023{\natexlab{a}}, Submitted to ApJ Supplement, arXiv:2310.12340

\bibitem[{{Eisenstein} {et~al.}(2023{\natexlab{b}}){Eisenstein}, {Willott}, {Alberts}, {Arribas}, {Bonaventura}, {Bunker}, {Cameron}, {Carniani}, {Charlot}, {Curtis-Lake}, {D'Eugenio}, {Endsley}, {Ferruit}, {Giardino}, {Hainline}, {Hausen}, {Jakobsen}, {Johnson}, {Maiolino}, {Rieke}, {Rieke}, {Rix}, {Robertson}, {Stark}, {Tacchella}, {Williams}, {Willmer}, {Baker}, {Baum}, {Bhatawdekar}, {Boyett}, {Chen}, {Chevallard}, {Circosta}, {Curti}, {Danhaive}, {DeCoursey}, {de Graaff}, {Dressler}, {Egami}, {Helton}, {Hviding}, {Ji}, {Jones}, {Kumari}, {L{\"u}tzgendorf}, {Laseter}, {Looser}, {Lyu}, {Maseda}, {Nelson}, {Parlanti}, {Perna}, {Pusk{\'a}s}, {Rawle}, {Rodr{\'\i}guez Del Pino}, {Sandles}, {Saxena}, {Scholtz}, {Sharpe}, {Shivaei}, {Silcock}, {Simmonds}, {Skarbinski}, {Smit}, {Stone}, {Suess}, {Sun}, {Tang}, {Topping}, {{\"U}bler}, {Villanueva}, {Wallace}, {Whitler}, {Witstok}, \& {Woodrum}}]{jades2}
{Eisenstein}, D.~J., {Willott}, C., {Alberts}, S., {et~al.} 2023{\natexlab{b}}, Submitted to ApJ Supplement, arXiv:2306.02465

\bibitem[{{Fagotto} {et~al.}(1994{\natexlab{a}}){Fagotto}, {Bressan}, {Bertelli}, \& {Chiosi}}]{Fagotto_1994a}
{Fagotto}, F., {Bressan}, A., {Bertelli}, G., \& {Chiosi}, C. 1994{\natexlab{a}}, \aaps, 104, 365

\bibitem[{{Fagotto} {et~al.}(1994{\natexlab{b}}){Fagotto}, {Bressan}, {Bertelli}, \& {Chiosi}}]{Fagotto_1994b}
{Fagotto}, F., {Bressan}, A., {Bertelli}, G., \& {Chiosi}, C. 1994{\natexlab{b}}, \aaps, 105, 29

\bibitem[{Fazio {et~al.}(2004)Fazio, Hora, Allen, Ashby, Barmby, Deutsch, Huang, Kleiner, Marengo, Megeath, Melnick, Pahre, Patten, Polizotti, Smith, Taylor, Wang, Willner, Hoffmann, Pipher, Forrest, McMurty, McCreight, McKelvey, McMurray, Koch, Moseley, Arendt, Mentzell, Marx, Losch, Mayman, Eichhorn, Krebs, Jhabvala, Gezari, Fixsen, Flores, Shakoorzadeh, Jungo, Hakun, Workman, Karpati, Kichak, Whitley, Mann, Tollestrup, Eisenhardt, Stern, Gorjian, Bhattacharya, Carey, Nelson, Glaccum, Lacy, Lowrance, Laine, Reach, Stauffer, Surace, Wilson, Wright, Hoffman, Domingo, \& Cohen}]{irac}
Fazio, G.~G., Hora, J.~L., Allen, L.~E., {et~al.} 2004, The Astrophysical Journal Supplement Series, 154, 10

\bibitem[{{Feuillet} {et~al.}(2024){Feuillet}, {Mel{\'e}ndez}, {Kraemer}, {Schmitt}, {Fischer}, \& {Reeves}}]{flr}
{Feuillet}, L.~M., {Mel{\'e}ndez}, M., {Kraemer}, S., {et~al.} 2024, \apj, 962, 104

\bibitem[{{Fukugita} {et~al.}(1996){Fukugita}, {Ichikawa}, {Gunn}, {Doi}, {Shimasaku}, \& {Schneider}}]{sdss_technical1}
{Fukugita}, M., {Ichikawa}, T., {Gunn}, J.~E., {et~al.} 1996, \aj, 111, 1748

\bibitem[{Gardner {et~al.}(2006)Gardner, Mather, Clampin, Doyon, Greenhouse, Hammel, Hutchings, Jakobsen, Lilly, Long, Lunine, Mccaughrean, Mountain, Nella, Rieke, Rieke, Rix, Smith, Sonneborn, Stiavelli, Stockman, Windhorst, \& Wright}]{jwst}
Gardner, J.~P., Mather, J.~C., Clampin, M., {et~al.} 2006, Space Science Reviews, 123, 485

\bibitem[{{Garilli} {et~al.}(2021){Garilli}, {McLure}, {Pentericci}, {Franzetti}, {Gargiulo}, {Carnall}, {Cucciati}, {Iovino}, {Amorin}, {Bolzonella}, {Bongiorno}, {Castellano}, {Cimatti}, {Cirasuolo}, {Cullen}, {Dunlop}, {Elbaz}, {Finkelstein}, {Fontana}, {Fontanot}, {Fumana}, {Guaita}, {Hartley}, {Jarvis}, {Juneau}, {Maccagni}, {McLeod}, {Nandra}, {Pompei}, {Pozzetti}, {Scodeggio}, {Talia}, {Calabr{\`o}}, {Cresci}, {Fynbo}, {Hathi}, {Hibon}, {Koekemoer}, {Magliocchetti}, {Salvato}, {Vietri}, {Zamorani}, {Almaini}, {Balestra}, {Bardelli}, {Begley}, {Brammer}, {Bell}, {Bowler}, {Brusa}, {Buitrago}, {Caputi}, {Cassata}, {Charlot}, {Citro}, {Cristiani}, {Curtis-Lake}, {Dickinson}, {Fazio}, {Ferguson}, {Fiore}, {Franco}, {Georgakakis}, {Giavalisco}, {Grazian}, {Hamadouche}, {Jung}, {Kim}, {Khusanova}, {Le F{\`e}vre}, {Longhetti}, {Lotz}, {Mannucci}, {Maltby}, {Matsuoka}, {Mendez-Hernandez}, {Mendez-Abreu}, {Mignoli}, {Moresco}, {Nonino}, {Pannella}, {Papovich}, {Popesso}, {Roberts-Borsani}, {Rosario},
  {Saldana-Lopez}, {Santini}, {Saxena}, {Schaerer}, {Schreiber}, {Stark}, {Tasca}, {Thomas}, {Vanzella}, {Wild}, {Williams}, \& {Zucca}}]{vandels_cat}
{Garilli}, B., {McLure}, R., {Pentericci}, L., {et~al.} 2021, \aap, 647, A150

\bibitem[{{Girardi} {et~al.}(1996){Girardi}, {Bressan}, {Chiosi}, {Bertelli}, \& {Nasi}}]{Girardi_1996}
{Girardi}, L., {Bressan}, A., {Chiosi}, C., {Bertelli}, G., \& {Nasi}, E. 1996, \aaps, 117, 113

\bibitem[{{Gomes} \& {Papaderos}(2017)}]{fado}
{Gomes}, J.~M. \& {Papaderos}, P. 2017, \aap, 603, A63

\bibitem[{{Groves} {et~al.}(2006){Groves}, {Heckman}, \& {Kauffmann}}]{groves06}
{Groves}, B.~A., {Heckman}, T.~M., \& {Kauffmann}, G. 2006, \mnras, 371, 1559

\bibitem[{{Gunn} {et~al.}(1998){Gunn}, {Carr}, {Rockosi}, {Sekiguchi}, {Berry}, {Elms}, {de Haas}, {Ivezi{\'c}}, {Knapp}, {Lupton}, {Pauls}, {Simcoe}, {Hirsch}, {Sanford}, {Wang}, {York}, {Harris}, {Annis}, {Bartozek}, {Boroski}, {Bakken}, {Haldeman}, {Kent}, {Holm}, {Holmgren}, {Petravick}, {Prosapio}, {Rechenmacher}, {Doi}, {Fukugita}, {Shimasaku}, {Okada}, {Hull}, {Siegmund}, {Mannery}, {Blouke}, {Heidtman}, {Schneider}, {Lucinio}, \& {Brinkman}}]{sdss_technical2}
{Gunn}, J.~E., {Carr}, M., {Rockosi}, C., {et~al.} 1998, \aj, 116, 3040

\bibitem[{{Hainline} {et~al.}(2024){Hainline}, {Johnson}, {Robertson}, {Tacchella}, {Helton}, {Sun}, {Eisenstein}, {Simmonds}, {Topping}, {Whitler}, {Willmer}, {Rieke}, {Suess}, {Hviding}, {Cameron}, {Alberts}, {Baker}, {Baum}, {Bhatawdekar}, {Bonaventura}, {Boyett}, {Bunker}, {Carniani}, {Charlot}, {Chevallard}, {Chen}, {Curti}, {Curtis-Lake}, {D'Eugenio}, {Egami}, {Endsley}, {Hausen}, {Ji}, {Looser}, {Lyu}, {Maiolino}, {Nelson}, {Pusk{\'a}s}, {Rawle}, {Sandles}, {Saxena}, {Smit}, {Stark}, {Williams}, {Willott}, \& {Witstok}}]{jades3}
{Hainline}, K.~N., {Johnson}, B.~D., {Robertson}, B., {et~al.} 2024, \apj, 964, 71

\bibitem[{{Hambly} {et~al.}(2008){Hambly}, {Collins}, {Cross}, {Mann}, {Read}, {Sutorius}, {Bond}, {Bryant}, {Emerson}, {Lawrence}, {Rimoldini}, {Stewart}, {Williams}, {Adamson}, {Hirst}, {Dye}, \& {Warren}}]{hambly08}
{Hambly}, N.~C., {Collins}, R.~S., {Cross}, N.~J.~G., {et~al.} 2008, \mnras, 384, 637

\bibitem[{{Harish} {et~al.}(2023){Harish}, {Malhotra}, {Rhoads}, {Jiang}, {Yang}, \& {Knorr}}]{harish23}
{Harish}, S., {Malhotra}, S., {Rhoads}, J.~E., {et~al.} 2023, \apj, 945, 157

\bibitem[{{Hewett} {et~al.}(2006){Hewett}, {Warren}, {Leggett}, \& {Hodgkin}}]{hewett06}
{Hewett}, P.~C., {Warren}, S.~J., {Leggett}, S.~K., \& {Hodgkin}, S.~T. 2006, \mnras, 367, 454

\bibitem[{{Hodgkin} {et~al.}(2009){Hodgkin}, {Irwin}, {Hewett}, \& {Warren}}]{hodgkin09}
{Hodgkin}, S.~T., {Irwin}, M.~J., {Hewett}, P.~C., \& {Warren}, S.~J. 2009, \mnras, 394, 675

\bibitem[{Hunter(2007)}]{matplotlib}
Hunter, J.~D. 2007, Computing in Science \& Engineering, 9, 90

\bibitem[{{Irwin} {et~al.}(2004){Irwin}, {Lewis}, {Hodgkin}, {Bunclark}, {Evans}, {McMahon}, {Emerson}, {Stewart}, \& {Beard}}]{irwin04}
{Irwin}, M.~J., {Lewis}, J., {Hodgkin}, S., {et~al.} 2004, in Society of Photo-Optical Instrumentation Engineers (SPIE) Conference Series, Vol. 5493, Optimizing Scientific Return for Astronomy through Information Technologies, ed. P.~J. {Quinn} \& A.~{Bridger}, 411--422

\bibitem[{{Juneau} {et~al.}(2011){Juneau}, {Dickinson}, {Alexander}, \& {Salim}}]{juneau11}
{Juneau}, S., {Dickinson}, M., {Alexander}, D.~M., \& {Salim}, S. 2011, \apj, 736, 104

\bibitem[{{Kauffmann} {et~al.}(2003){Kauffmann}, {Heckman}, {Tremonti}, {Brinchmann}, {Charlot}, {White}, {Ridgway}, {Brinkmann}, {Fukugita}, {Hall}, {Ivezi{\'c}}, {Richards}, \& {Schneider}}]{kauffmann_2003}
{Kauffmann}, G., {Heckman}, T.~M., {Tremonti}, C., {et~al.} 2003, \mnras, 346, 1055

\bibitem[{{Kewley} \& {Dopita}(2002)}]{kewley_2002}
{Kewley}, L.~J. \& {Dopita}, M.~A. 2002, \apjs, 142, 35

\bibitem[{{Kewley} {et~al.}(2001){Kewley}, {Dopita}, {Sutherland}, {Heisler}, \& {Trevena}}]{kewley_2001}
{Kewley}, L.~J., {Dopita}, M.~A., {Sutherland}, R.~S., {Heisler}, C.~A., \& {Trevena}, J. 2001, \apj, 556, 121

\bibitem[{{Kewley} {et~al.}(2006){Kewley}, {Groves}, {Kauffmann}, \& {Heckman}}]{kewley}
{Kewley}, L.~J., {Groves}, B., {Kauffmann}, G., \& {Heckman}, T. 2006, \mnras, 372, 961

\bibitem[{Kewley {et~al.}(2013)Kewley, Maier, Yabe, Ohta, Akiyama, Dopita, \& Yuan}]{kewley_2013}
Kewley, L.~J., Maier, C., Yabe, K., {et~al.} 2013, The Astrophysical Journal, 774, L10

\bibitem[{{Khostovan} {et~al.}(2016){Khostovan}, {Sobral}, {Mobasher}, {Smail}, {Darvish}, {Nayyeri}, {Hemmati}, \& {Stott}}]{khostovan2016}
{Khostovan}, A.~A., {Sobral}, D., {Mobasher}, B., {et~al.} 2016, \mnras, 463, 2363

\bibitem[{{Kriek} {et~al.}(2015){Kriek}, {Shapley}, {Reddy}, {Siana}, {Coil}, {Mobasher}, {Freeman}, {de Groot}, {Price}, {Sanders}, {Shivaei}, {Brammer}, {Momcheva}, {Skelton}, {van Dokkum}, {Whitaker}, {Aird}, {Azadi}, {Kassis}, {Bullock}, {Conroy}, {Dav{\'e}}, {Kere{\v{s}}}, \& {Krumholz}}]{kriek_15}
{Kriek}, M., {Shapley}, A.~E., {Reddy}, N.~A., {et~al.} 2015, \apjs, 218, 15

\bibitem[{{Lagos} {et~al.}(2022){Lagos}, {Loubser}, {Scott}, {O'Sullivan}, {Kolokythas}, {Babul}, {Nigoche-Netro}, {Olivares}, \& {Sengupta}}]{patricio}
{Lagos}, P., {Loubser}, S.~I., {Scott}, T.~C., {et~al.} 2022, \mnras, 516, 5487

\bibitem[{{Lam} {et~al.}(2025){Lam}, {Shapley}, {Sanders}, {Do}, {Jones}, {Coil}, {Kriek}, {Mobasher}, {Reddy}, {Siana}, \& {Clarke}}]{lam25}
{Lam}, N., {Shapley}, A.~E., {Sanders}, R.~L., {et~al.} 2025, Submitted to ApJ, arXiv:2506.22547

\bibitem[{LeFevre {et~al.}(2003)LeFevre, Saisse, Mancini, Brau-Nogue, Caputi, Castinel, D'Odorico, Garilli, Kissler-Patig, Lucuix, Mancini, Pauget, Sciarretta, Scodeggio, Tresse, \& Vettolani}]{vimos}
LeFevre, O., Saisse, M., Mancini, D., {et~al.} 2003, in Instrument Design and Performance for Optical/Infrared Ground-based Telescopes, ed. M.~Iye \& A.~F.~M. Moorwood, Vol. 4841, International Society for Optics and Photonics (SPIE), 1670 -- 1681

\bibitem[{{Madau} \& {Dickinson}(2014)}]{madaudickinson_14}
{Madau}, P. \& {Dickinson}, M. 2014, \araa, 52, 415

\bibitem[{{Maiolino} {et~al.}(2020){Maiolino}, {Cirasuolo}, {Afonso}, {Bauer}, {Bowler}, {Cucciati}, {Daddi}, {De Lucia}, {Evans}, {Flores}, {Gargiulo}, {Garilli}, {Jablonka}, {Jarvis}, {Kneib}, {Lilly}, {Looser}, {Magliocchetti}, {Man}, {Mannucci}, {Maurogordato}, {McLure}, {Norberg}, {Oesch}, {Oliva}, {Paltani}, {Pappalardo}, {Peng}, {Pentericci}, {Pozzetti}, {Renzini}, {Rodrigues}, {Royer}, {Serjeant}, {Vanzi}, {Wild}, \& {Zamorani}}]{moon3}
{Maiolino}, R., {Cirasuolo}, M., {Afonso}, J., {et~al.} 2020, The Messenger, 180, 24

\bibitem[{{Maiolino} {et~al.}(2008){Maiolino}, {Nagao}, {Grazian}, {Cocchia}, {Marconi}, {Mannucci}, {Cimatti}, {Pipino}, {Ballero}, {Calura}, {Chiappini}, {Fontana}, {Granato}, {Matteucci}, {Pastorini}, {Pentericci}, {Risaliti}, {Salvati}, \& {Silva}}]{maiolino08}
{Maiolino}, R., {Nagao}, T., {Grazian}, A., {et~al.} 2008, \aap, 488, 463

\bibitem[{{Mazzolari} {et~al.}(2024){Mazzolari}, {{\"U}bler}, {Maiolino}, {Ji}, {Nakajima}, {Feltre}, {Scholtz}, {D'Eugenio}, {Curti}, {Mignoli}, \& {Marconi}}]{mazzolari24}
{Mazzolari}, G., {{\"U}bler}, H., {Maiolino}, R., {et~al.} 2024, \aap, 691, A345

\bibitem[{{McLean} {et~al.}(2012){McLean}, {Steidel}, {Epps}, {Konidaris}, {Matthews}, {Adkins}, {Aliado}, {Brims}, {Canfield}, {Cromer}, {Fucik}, {Kulas}, {Mace}, {Magnone}, {Rodriguez}, {Rudie}, {Trainor}, {Wang}, {Weber}, \& {Weiss}}]{mosfire}
{McLean}, I.~S., {Steidel}, C.~C., {Epps}, H.~W., {et~al.} 2012, in Society of Photo-Optical Instrumentation Engineers (SPIE) Conference Series, Vol. 8446, Ground-based and Airborne Instrumentation for Astronomy IV, ed. I.~S. {McLean}, S.~K. {Ramsay}, \& H.~{Takami}, 84460J

\bibitem[{{McLure} {et~al.}(2018){McLure}, {Pentericci}, {Cimatti}, {Dunlop}, {Elbaz}, {Fontana}, {Nandra}, {Amorin}, {Bolzonella}, {Bongiorno}, {Carnall}, {Castellano}, {Cirasuolo}, {Cucciati}, {Cullen}, {De Barros}, {Finkelstein}, {Fontanot}, {Franzetti}, {Fumana}, {Gargiulo}, {Garilli}, {Guaita}, {Hartley}, {Iovino}, {Jarvis}, {Juneau}, {Karman}, {Maccagni}, {Marchi}, {M{\'a}rmol-Queralt{\'o}}, {Pompei}, {Pozzetti}, {Scodeggio}, {Sommariva}, {Talia}, {Almaini}, {Balestra}, {Bardelli}, {Bell}, {Bourne}, {Bowler}, {Brusa}, {Buitrago}, {Caputi}, {Cassata}, {Charlot}, {Citro}, {Cresci}, {Cristiani}, {Curtis-Lake}, {Dickinson}, {Fazio}, {Ferguson}, {Fiore}, {Franco}, {Fynbo}, {Galametz}, {Georgakakis}, {Giavalisco}, {Grazian}, {Hathi}, {Jung}, {Kim}, {Koekemoer}, {Khusanova}, {Le F{\`e}vre}, {Lotz}, {Mannucci}, {Maltby}, {Matsuoka}, {McLeod}, {Mendez-Hernandez}, {Mendez-Abreu}, {Mignoli}, {Moresco}, {Mortlock}, {Nonino}, {Pannella}, {Papovich}, {Popesso}, {Rosario}, {Salvato}, {Santini}, {Schaerer},
  {Schreiber}, {Stark}, {Tasca}, {Thomas}, {Treu}, {Vanzella}, {Wild}, {Williams}, {Zamorani}, \& {Zucca}}]{vandels2}
{McLure}, R.~J., {Pentericci}, L., {Cimatti}, A., {et~al.} 2018, \mnras, 479, 25

\bibitem[{{Miranda} {et~al.}(2023){Miranda}, {Pappalardo}, {Papaderos}, {Afonso}, {Matute}, {Lobo}, {Paulino-Afonso}, {Carvajal}, {Lorenzoni}, \& {Santos}}]{henrique}
{Miranda}, H., {Pappalardo}, C., {Papaderos}, P., {et~al.} 2023, \aap, 669, A16

\bibitem[{{Momcheva} {et~al.}(2016){Momcheva}, {Brammer}, {van Dokkum}, {Skelton}, {Whitaker}, {Nelson}, {Fumagalli}, {Maseda}, {Leja}, {Franx}, {Rix}, {Bezanson}, {Da Cunha}, {Dickey}, {F{\"o}rster Schreiber}, {Illingworth}, {Kriek}, {Labb{\'e}}, {Ulf Lange}, {Lundgren}, {Magee}, {Marchesini}, {Oesch}, {Pacifici}, {Patel}, {Price}, {Tal}, {Wake}, {van der Wel}, \& {Wuyts}}]{momcheva_16}
{Momcheva}, I.~G., {Brammer}, G.~B., {van Dokkum}, P.~G., {et~al.} 2016, \apjs, 225, 27

\bibitem[{{Nakajima} {et~al.}(2022){Nakajima}, {Ouchi}, {Xu}, {Rauch}, {Harikane}, {Nishigaki}, {Isobe}, {Kusakabe}, {Nagao}, {Ono}, {Onodera}, {Sugahara}, {Kim}, {Komiyama}, {Lee}, \& {Zahedy}}]{nakajima22}
{Nakajima}, K., {Ouchi}, M., {Xu}, Y., {et~al.} 2022, \apjs, 262, 3

\bibitem[{Osterbrock \& Ferland(2006)}]{osterbrock}
Osterbrock, D. \& Ferland, G. 2006, Astrophysics Of Gas Nebulae and Active Galactic Nuclei (University Science Books)

\bibitem[{{Paalvast} {et~al.}(2018){Paalvast}, {Verhamme}, {Straka}, {Brinchmann}, {Herenz}, {Carton}, {Gunawardhana}, {Boogaard}, {Cantalupo}, {Contini}, {Epinat}, {Inami}, {Marino}, {Maseda}, {Michel-Dansac}, {Muzahid}, {Nanayakkara}, {Pezzulli}, {Richard}, {Schaye}, {Segers}, {Urrutia}, {Wendt}, \& {Wisotzki}}]{paalvast}
{Paalvast}, M., {Verhamme}, A., {Straka}, L.~A., {et~al.} 2018, \aap, 618, A40

\bibitem[{{Papaderos} {et~al.}(2013){Papaderos}, {Gomes}, {V{\'\i}lchez}, {Kehrig}, {Lehnert}, {Ziegler}, {S{\'a}nchez}, {Husemann}, {Monreal-Ibero}, {Garc{\'\i}a-Benito}, {Bland-Hawthorn}, {Cortijo-Ferrero}, {de Lorenzo-C{\'a}ceres}, {del Olmo}, {Falc{\'o}n-Barroso}, {Galbany}, {Iglesias-P{\'a}ramo}, {L{\'o}pez-S{\'a}nchez}, {Marquez}, {Moll{\'a}}, {Mast}, {van de Ven}, \& {Wisotzki}}]{papaderos2013}
{Papaderos}, P., {Gomes}, J.~M., {V{\'\i}lchez}, J.~M., {et~al.} 2013, \aap, 555, L1

\bibitem[{{Papaderos} {et~al.}(2023){Papaderos}, {{\"O}stlin}, \& {Breda}}]{papaderos23}
{Papaderos}, P., {{\"O}stlin}, G., \& {Breda}, I. 2023, \aap, 673, A30

\bibitem[{{Pentericci} {et~al.}(2018){Pentericci}, {McLure}, {Garilli}, {Cucciati}, {Franzetti}, {Iovino}, {Amorin}, {Bolzonella}, {Bongiorno}, {Carnall}, {Castellano}, {Cimatti}, {Cirasuolo}, {Cullen}, {De Barros}, {Dunlop}, {Elbaz}, {Finkelstein}, {Fontana}, {Fontanot}, {Fumana}, {Gargiulo}, {Guaita}, {Hartley}, {Jarvis}, {Juneau}, {Karman}, {Maccagni}, {Marchi}, {Marmol-Queralto}, {Nandra}, {Pompei}, {Pozzetti}, {Scodeggio}, {Sommariva}, {Talia}, {Almaini}, {Balestra}, {Bardelli}, {Bell}, {Bourne}, {Bowler}, {Brusa}, {Buitrago}, {Caputi}, {Cassata}, {Charlot}, {Citro}, {Cresci}, {Cristiani}, {Curtis-Lake}, {Dickinson}, {Fazio}, {Ferguson}, {Fiore}, {Franco}, {Fynbo}, {Galametz}, {Georgakakis}, {Giavalisco}, {Grazian}, {Hathi}, {Jung}, {Kim}, {Koekemoer}, {Khusanova}, {Le F{\`e}vre}, {Lotz}, {Mannucci}, {Maltby}, {Matsuoka}, {McLeod}, {Mendez-Hernandez}, {Mendez-Abreu}, {Mignoli}, {Moresco}, {Mortlock}, {Nonino}, {Pannella}, {Papovich}, {Popesso}, {Rosario}, {Salvato}, {Santini}, {Schaerer}, {Schreiber},
  {Stark}, {Tasca}, {Thomas}, {Treu}, {Vanzella}, {Wild}, {Williams}, {Zamorani}, \& {Zucca}}]{vandels1}
{Pentericci}, L., {McLure}, R.~J., {Garilli}, B., {et~al.} 2018, \aap, 616, A174

\bibitem[{{Perrotta} {et~al.}(2021){Perrotta}, {George}, {Coil}, {Tremonti}, {Rupke}, {Davis}, {Diamond-Stanic}, {Geach}, {Hickox}, {Moustakas}, {Petter}, {Rudnick}, {Sell}, {Swiggum}, \& {Whalen}}]{perrotta}
{Perrotta}, S., {George}, E.~R., {Coil}, A.~L., {et~al.} 2021, \apj, 923, 275

\bibitem[{{Polimera} {et~al.}(2022){Polimera}, {Kannappan}, {Richardson}, {Bittner}, {Ferguson}, {Moffett}, {Eckert}, {Bellovary}, \& {Norris}}]{polimera22}
{Polimera}, M.~S., {Kannappan}, S.~J., {Richardson}, C.~T., {et~al.} 2022, \apj, 931, 44

\bibitem[{{Reddy} {et~al.}(2015){Reddy}, {Kriek}, {Shapley}, {Freeman}, {Siana}, {Coil}, {Mobasher}, {Price}, {Sanders}, \& {Shivaei}}]{reddy_15}
{Reddy}, N.~A., {Kriek}, M., {Shapley}, A.~E., {et~al.} 2015, \apj, 806, 259

\bibitem[{{Rieke} {et~al.}(2023){Rieke}, {Robertson}, {Tacchella}, {Hainline}, {Johnson}, {Hausen}, {Ji}, {Willmer}, {Eisenstein}, {Pusk{\'a}s}, {Alberts}, {Arribas}, {Baker}, {Baum}, {Bhatawdekar}, {Bonaventura}, {Boyett}, {Bunker}, {Cameron}, {Carniani}, {Charlot}, {Chevallard}, {Chen}, {Curti}, {Curtis-Lake}, {Danhaive}, {DeCoursey}, {Dressler}, {Egami}, {Endsley}, {Helton}, {Hviding}, {Kumari}, {Looser}, {Lyu}, {Maiolino}, {Maseda}, {Nelson}, {Rieke}, {Rix}, {Sandles}, {Saxena}, {Sharpe}, {Shivaei}, {Skarbinski}, {Smit}, {Stark}, {Stone}, {Suess}, {Sun}, {Topping}, {{\"U}bler}, {Villanueva}, {Wallace}, {Williams}, {Willott}, {Whitler}, {Witstok}, \& {Woodrum}}]{jades4}
{Rieke}, M.~J., {Robertson}, B., {Tacchella}, S., {et~al.} 2023, \apjs, 269, 16

\bibitem[{Rola {et~al.}(1997)Rola, Terlevich, \& Terlevich}]{Rola97}
Rola, C.~S., Terlevich, E., \& Terlevich, R.~J. 1997, Monthly Notices of the Royal Astronomical Society, 289, 419

\bibitem[{{Salpeter}(1955)}]{salpeter}
{Salpeter}, E.~E. 1955, \apj, 121, 161

\bibitem[{{Schawinski} {et~al.}(2007){Schawinski}, {Thomas}, {Sarzi}, {Maraston}, {Kaviraj}, {Joo}, {Yi}, \& {Silk}}]{schaw}
{Schawinski}, K., {Thomas}, D., {Sarzi}, M., {et~al.} 2007, \mnras, 382, 1415

\bibitem[{{Schlafly} \& {Finkbeiner}(2011)}]{Schlafly_2011}
{Schlafly}, E.~F. \& {Finkbeiner}, D.~P. 2011, \apj, 737, 103

\bibitem[{{Schlegel} {et~al.}(1998){Schlegel}, {Finkbeiner}, \& {Davis}}]{Schlegel_1998}
{Schlegel}, D.~J., {Finkbeiner}, D.~P., \& {Davis}, M. 1998, \apj, 500, 525

\bibitem[{{Scoville} {et~al.}(2007){Scoville}, {Aussel}, {Brusa}, {Capak}, {Carollo}, {Elvis}, {Giavalisco}, {Guzzo}, {Hasinger}, {Impey}, {Kneib}, {LeFevre}, {Lilly}, {Mobasher}, {Renzini}, {Rich}, {Sanders}, {Schinnerer}, {Schminovich}, {Shopbell}, {Taniguchi}, \& {Tyson}}]{cosmos}
{Scoville}, N., {Aussel}, H., {Brusa}, M., {et~al.} 2007, \apjs, 172, 1

\bibitem[{{Searle}(1972)}]{searle72}
{Searle}, L. 1972, in External Galaxies and Quasi-Stellar Objects, ed. D.~S. {Evans}, D.~{Wills}, \& B.~J. {Wills}, Vol.~44, 66

\bibitem[{{Shapley} {et~al.}(2015){Shapley}, {Reddy}, {Kriek}, {Freeman}, {Sanders}, {Siana}, {Coil}, {Mobasher}, {Shivaei}, {Price}, \& {de Groot}}]{shapley_2015}
{Shapley}, A.~E., {Reddy}, N.~A., {Kriek}, M., {et~al.} 2015, \apj, 801, 88

\bibitem[{{Shapley} {et~al.}(2025){Shapley}, {Sanders}, {Topping}, {Reddy}, {Berg}, {Bouwens}, {Brammer}, {Carnall}, {Cullen}, {Dav{\'e}}, {Dunlop}, {Ellis}, {F{\"o}rster Schreiber}, {Furlanetto}, {Glazebrook}, {Illingworth}, {Jones}, {Kriek}, {McLeod}, {McLure}, {Narayanan}, {Oesch}, {Pahl}, {Pettini}, {Schaerer}, {Stark}, {Steidel}, {Tang}, {Clarke}, {Donnan}, \& {Kehoe}}]{shapley25}
{Shapley}, A.~E., {Sanders}, R.~L., {Topping}, M.~W., {et~al.} 2025, \apj, 980, 242

\bibitem[{{Skelton} {et~al.}(2014){Skelton}, {Whitaker}, {Momcheva}, {Brammer}, {van Dokkum}, {Labb{\'e}}, {Franx}, {van der Wel}, {Bezanson}, {Da Cunha}, {Fumagalli}, {F{\"o}rster Schreiber}, {Kriek}, {Leja}, {Lundgren}, {Magee}, {Marchesini}, {Maseda}, {Nelson}, {Oesch}, {Pacifici}, {Patel}, {Price}, {Rix}, {Tal}, {Wake}, \& {Wuyts}}]{skelton14}
{Skelton}, R.~E., {Whitaker}, K.~E., {Momcheva}, I.~G., {et~al.} 2014, \apjs, 214, 24

\bibitem[{{Solimano} {et~al.}(2025){Solimano}, {Gonz{\'a}lez-L{\'o}pez}, {Aravena}, {Alcalde Pampliega}, {Assef}, {B{\'e}thermin}, {Boquien}, {Bovino}, {Casey}, {Cassata}, {da Cunha}, {Davies}, {De Looze}, {Ding}, {D{\'\i}az-Santos}, {Faisst}, {Ferrara}, {Fisher}, {F{\"o}rster-Schreiber}, {Fujimoto}, {Ginolfi}, {Gruppioni}, {Guaita}, {Hathi}, {Herrera-Camus}, {Ibar}, {Inami}, {Jones}, {Koekemoer}, {Lee}, {Li}, {Liu}, {Liu}, {Molina}, {Ogle}, {Posses}, {Pozzi}, {Rela{\~n}o}, {Riechers}, {Romano}, {Spilker}, {Sulzenauer}, {Telikova}, {Vallini}, {Vasan}, {Veilleux}, {Vergani}, {Villanueva}, {Wang}, {Yan}, \& {Zamorani}}]{heii}
{Solimano}, M., {Gonz{\'a}lez-L{\'o}pez}, J., {Aravena}, M., {et~al.} 2025, \aap, 693, A70

\bibitem[{{Stasi{\'n}ska} {et~al.}(2015){Stasi{\'n}ska}, {Izotov}, {Morisset}, \& {Guseva}}]{stasinka_2018}
{Stasi{\'n}ska}, G., {Izotov}, Y., {Morisset}, C., \& {Guseva}, N. 2015, \aap, 576, A83

\bibitem[{{Stasi{\'n}ska} {et~al.}(2001){Stasi{\'n}ska}, {Schaerer}, \& {Leitherer}}]{stasinska01}
{Stasi{\'n}ska}, G., {Schaerer}, D., \& {Leitherer}, C. 2001, \aap, 370, 1

\bibitem[{{Steidel} {et~al.}(2014){Steidel}, {Rudie}, {Strom}, {Pettini}, {Reddy}, {Shapley}, {Trainor}, {Erb}, {Turner}, {Konidaris}, {Kulas}, {Mace}, {Matthews}, \& {McLean}}]{steidel_2014}
{Steidel}, C.~C., {Rudie}, G.~C., {Strom}, A.~L., {et~al.} 2014, \apj, 795, 165

\bibitem[{{Steidel} {et~al.}(2016){Steidel}, {Strom}, {Pettini}, {Rudie}, {Reddy}, \& {Trainor}}]{steidel_2016}
{Steidel}, C.~C., {Strom}, A.~L., {Pettini}, M., {et~al.} 2016, \apj, 826, 159

\bibitem[{{Straatman} {et~al.}(2018){Straatman}, {van der Wel}, {Bezanson}, {Pacifici}, {Gallazzi}, {Wu}, {Noeske}, {Bari{\v{s}}i{\'c}}, {Bell}, {Brammer}, {Calhau}, {Chauke}, {Franx}, {van Houdt}, {Labb{\'e}}, {Maseda}, {Mu{\~n}oz-Mateos}, {Muzzin}, {van de Sande}, {Sobral}, \& {Spilker}}]{legac2}
{Straatman}, C. M.~S., {van der Wel}, A., {Bezanson}, R., {et~al.} 2018, \apjs, 239, 27

\bibitem[{{Szokoly} {et~al.}(2004){Szokoly}, {Bergeron}, {Hasinger}, {Lehmann}, {Kewley}, {Mainieri}, {Nonino}, {Rosati}, {Giacconi}, {Gilli}, {Gilmozzi}, {Norman}, {Romaniello}, {Schreier}, {Tozzi}, {Wang}, {Zheng}, \& {Zirm}}]{cdfs}
{Szokoly}, G.~P., {Bergeron}, J., {Hasinger}, G., {et~al.} 2004, \apjs, 155, 271

\bibitem[{{Takada} {et~al.}(2014){Takada}, {Ellis}, {Chiba}, {Greene}, {Aihara}, {Arimoto}, {Bundy}, {Cohen}, {Dor{\'e}}, {Graves}, {Gunn}, {Heckman}, {Hirata}, {Ho}, {Kneib}, {Le F{\`e}vre}, {Lin}, {More}, {Murayama}, {Nagao}, {Ouchi}, {Seiffert}, {Silverman}, {Sodr{\'e}}, {Spergel}, {Strauss}, {Sugai}, {Suto}, {Takami}, \& {Wyse}}]{pfs}
{Takada}, M., {Ellis}, R.~S., {Chiba}, M., {et~al.} 2014, \pasj, 66, R1

\bibitem[{Teimoorinia \& Keown(2018)}]{Teimoorinia18}
Teimoorinia, H. \& Keown, J. 2018, Monthly Notices of the Royal Astronomical Society, 478, 3177

\bibitem[{{Teimoorinia} {et~al.}(2024){Teimoorinia}, {Shishehchi}, {Archinuk}, {Woo}, {Bickley}, {Lin}, {Hu}, \& {Petit}}]{teimoorinia24}
{Teimoorinia}, H., {Shishehchi}, S., {Archinuk}, F., {et~al.} 2024, \apj, 973, 95

\bibitem[{{Tremonti} {et~al.}(2004){Tremonti}, {Heckman}, {Kauffmann}, {Brinchmann}, {Charlot}, {White}, {Seibert}, {Peng}, {Schlegel}, {Uomoto}, {Fukugita}, \& {Brinkmann}}]{Tremonti_2004}
{Tremonti}, C.~A., {Heckman}, T.~M., {Kauffmann}, G., {et~al.} 2004, \apj, 613, 898

\bibitem[{{van der Wel} {et~al.}(2021){van der Wel}, {Bezanson}, {D'Eugenio}, {Straatman}, {Franx}, {van Houdt}, {Maseda}, {Gallazzi}, {Wu}, {Pacifici}, {Barisic}, {Brammer}, {Munoz-Mateos}, {Vervalcke}, {Zibetti}, {Sobral}, {de Graaff}, {Calhau}, {Kaushal}, {Muzzin}, {Bell}, \& {van Dokkum}}]{legac_cat}
{van der Wel}, A., {Bezanson}, R., {D'Eugenio}, F., {et~al.} 2021, \apjs, 256, 44

\bibitem[{{van der Wel} {et~al.}(2016){van der Wel}, {Noeske}, {Bezanson}, {Pacifici}, {Gallazzi}, {Franx}, {Mu{\~n}oz-Mateos}, {Bell}, {Brammer}, {Charlot}, {Chauk{\'e}}, {Labb{\'e}}, {Maseda}, {Muzzin}, {Rix}, {Sobral}, {van de Sande}, {van Dokkum}, {Wild}, \& {Wolf}}]{legac1}
{van der Wel}, A., {Noeske}, K., {Bezanson}, R., {et~al.} 2016, \apjs, 223, 29

\bibitem[{{Veilleux} \& {Osterbrock}(1987)}]{veilleux}
{Veilleux}, S. \& {Osterbrock}, D.~E. 1987, \apjs, 63, 295

\bibitem[{{Werner} {et~al.}(2004){Werner}, {Roellig}, {Low}, {Rieke}, {Rieke}, {Hoffmann}, {Young}, {Houck}, {Brandl}, {Fazio}, {Hora}, {Gehrz}, {Helou}, {Soifer}, {Stauffer}, {Keene}, {Eisenhardt}, {Gallagher}, {Gautier}, {Irace}, {Lawrence}, {Simmons}, {Van Cleve}, {Jura}, {Wright}, \& {Cruikshank}}]{Spitzer}
{Werner}, M.~W., {Roellig}, T.~L., {Low}, F.~J., {et~al.} 2004, \apjs, 154, 1

\bibitem[{{York} {et~al.}(2000){York}, {Adelman}, {Anderson}, {Anderson}, {Annis}, {Bahcall}, {Bakken}, {Barkhouser}, {Bastian}, {Berman}, {Boroski}, {Bracker}, {Briegel}, {Briggs}, {Brinkmann}, {Brunner}, {Burles}, {Carey}, {Carr}, {Castander}, {Chen}, {Colestock}, {Connolly}, {Crocker}, {Csabai}, {Czarapata}, {Davis}, {Doi}, {Dombeck}, {Eisenstein}, {Ellman}, {Elms}, {Evans}, {Fan}, {Federwitz}, {Fiscelli}, {Friedman}, {Frieman}, {Fukugita}, {Gillespie}, {Gunn}, {Gurbani}, {de Haas}, {Haldeman}, {Harris}, {Hayes}, {Heckman}, {Hennessy}, {Hindsley}, {Holm}, {Holmgren}, {Huang}, {Hull}, {Husby}, {Ichikawa}, {Ichikawa}, {Ivezi{\'c}}, {Kent}, {Kim}, {Kinney}, {Klaene}, {Kleinman}, {Kleinman}, {Knapp}, {Korienek}, {Kron}, {Kunszt}, {Lamb}, {Lee}, {Leger}, {Limmongkol}, {Lindenmeyer}, {Long}, {Loomis}, {Loveday}, {Lucinio}, {Lupton}, {MacKinnon}, {Mannery}, {Mantsch}, {Margon}, {McGehee}, {McKay}, {Meiksin}, {Merelli}, {Monet}, {Munn}, {Narayanan}, {Nash}, {Neilsen}, {Neswold}, {Newberg}, {Nichol}, {Nicinski},
  {Nonino}, {Okada}, {Okamura}, {Ostriker}, {Owen}, {Pauls}, {Peoples}, {Peterson}, {Petravick}, {Pier}, {Pope}, {Pordes}, {Prosapio}, {Rechenmacher}, {Quinn}, {Richards}, {Richmond}, {Rivetta}, {Rockosi}, {Ruthmansdorfer}, {Sandford}, {Schlegel}, {Schneider}, {Sekiguchi}, {Sergey}, {Shimasaku}, {Siegmund}, {Smee}, {Smith}, {Snedden}, {Stone}, {Stoughton}, {Strauss}, {Stubbs}, {SubbaRao}, {Szalay}, {Szapudi}, {Szokoly}, {Thakar}, {Tremonti}, {Tucker}, {Uomoto}, {Vanden Berk}, {Vogeley}, {Waddell}, {Wang}, {Watanabe}, {Weinberg}, {Yanny}, {Yasuda}, \& {SDSS Collaboration}}]{sdss_technical3}
{York}, D.~G., {Adelman}, J., {Anderson}, John~E., J., {et~al.} 2000, \aj, 120, 1579

\end{thebibliography}

\begin{appendix}

\section{Higher redshift AGN Classification}\label{appendix:highz class}

In this appendix, we provide further information into how AGNs were estimated from the different datasets for our higher redshift sample, described in Sects. \ref{section:sample} and \ref{section:highz}. All the information on how many galaxies are classified as SF and AGN according to each dataset is summed in Table \ref{tab:appendix sfagn}.

\begin{table}[h]
\centering
\caption{SF galaxies and AGNs found by each higher redshift dataset considered in this work.}
\label{tab:appendix sfagn}
\begin{tabular}{c|ccc}
Classification & SF       & AGN & Total \\ \hline
LEGA-C         & 699      & 24  & 723    \\
VANDELS        & 3        & 0   & 3     \\
MOSDEF         & 67       & 3   & 70     \\
3D-HST         & 326      & 20  & 346    \\
FADO-JWST      & 10       & 0   & 10     \\ \hline
Total          & $1\,105$ & 47  & $1\,152$    \\ \hline
\end{tabular}
\end{table}

\subsection{LEGA-C}

\begin{figure}[h]
    \centering
    \includegraphics[width=1.0\linewidth]{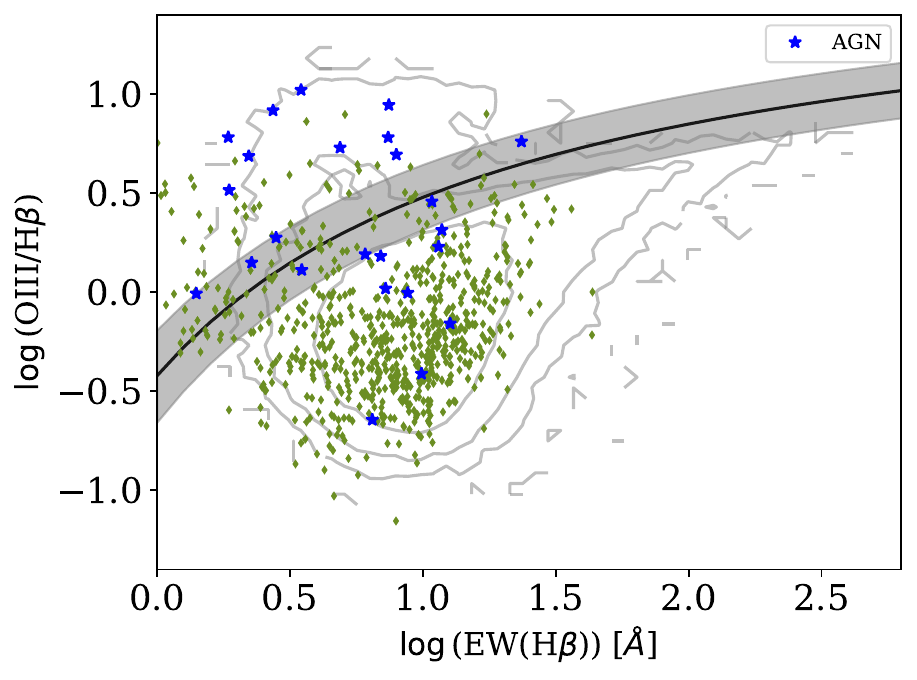}
    \caption{OB-I diagram with the LEGA-C sample. The blue stars represent AGNs according to the \texttt{flag\_spec} criterium. The green diamonds should represent SF galaxies. The grey contours represent the FADO-SDSS sample and the black line and grey shaded area represents our empirical separation given by Eq. \ref{eq:hbeta limit}.}
    \label{fig:appendix lega-c}
\end{figure}

In the LEGA-C sample, the \texttt{flag\_spec} column lets us know that there is an AGN present in the sample if this flag is equal to 1. These objects were flagged as AGNs due to clear evidence that the continuum was affected by this component and this, in turn, affected the measurement of indices, which was confirmed by visual inspection. With this scheme, we mark 24 galaxies as AGNs, and the remaining 699 objects are likely to be SF. This sample can be seen on the OB-I diagram in Fig. \ref{fig:appendix lega-c}.

\subsection{3D-HST}

\begin{figure}[h]
    \centering
    \includegraphics[width=1.0\linewidth]{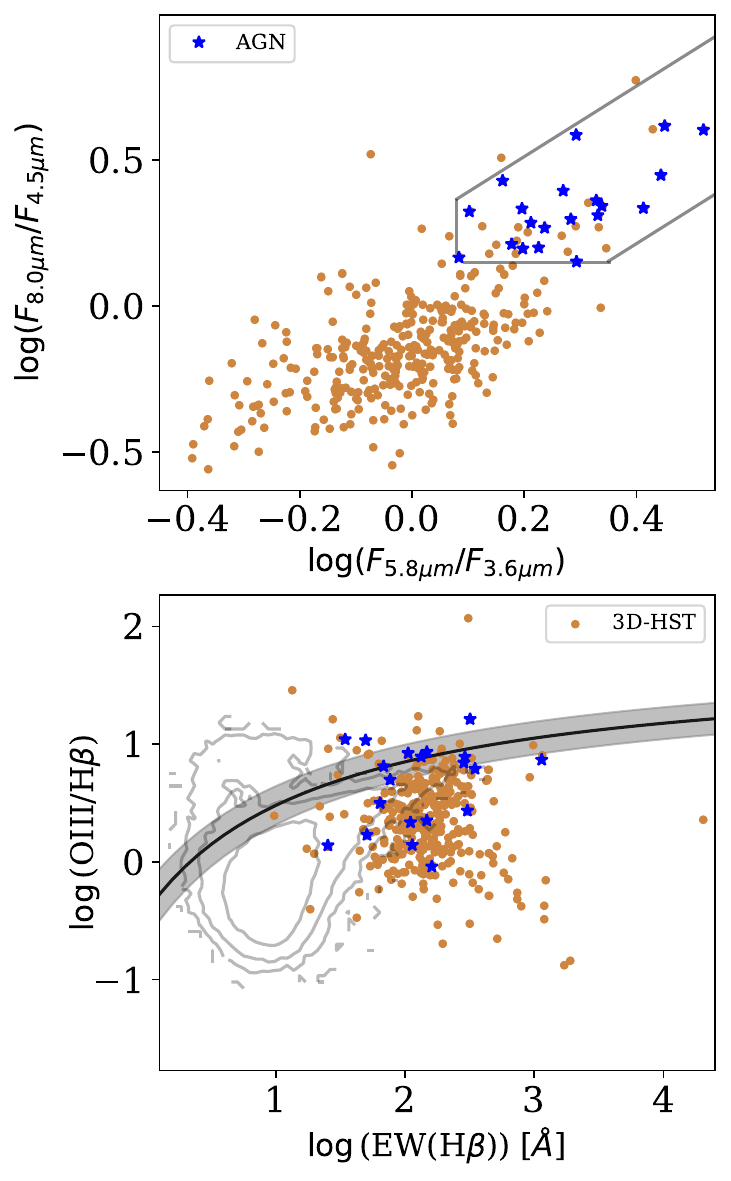}
    \caption{3D-HST sample in an IRAC colour-colour plot (top) and the OB-I diagram (bottom). \textit{Top:} The black lines represent the area according to Eq. \ref{eq:irac} that was used to identify AGNs (blue stars). All remaining galaxies should be SF and are represented by orange dots. \textit{Bottom:} The stars and dots represent the same as in the IRAC colour-colour plot. The grey contours represent the FADO-SDSS sample and the black line and grey shaded area represents our empirical separation given by Eq. \ref{eq:hbeta limit}.}
    \label{fig:appendix 3dhst}
\end{figure}

Regarding 3D-HST data, we can use the available IRAC photometry to select AGNs. Using the classification scheme by \cite{donley2012}, a galaxy is defined as an AGN if it obeys the following conditions:
\begin{flalign}
 \begin{aligned}
   &\log\left(\frac{F_{5.8 \mu m}}{F_{3.6 \mu m}}\right) \geqslant 0.08\\ & \log\left(\frac{F_{8.0 \mu m}}{F_{4.5 \mu m}}\right) \geqslant 0.15 \\ &
   \log\left(\frac{F_{8.0 \mu m}}{F_{4.5 \mu m}}\right) \geqslant 1.21 \; \log\left(\frac{F_{5.8 \mu m}}{F_{3.6 \mu m}}\right) - 0.27 \\ &
   \log\left(\frac{F_{8.0 \mu m}}{F_{4.5 \mu m}}\right) \leqslant 1.21 \; \log\left(\frac{F_{5.8 \mu m}}{F_{3.6 \mu m}}\right) + 0.27 \\ &
   F_{4.5 \mu m} > F_{3.6 \mu m} \\ &
   F_{5.8 \mu m} > F_{4.5 \mu m} \\ &
   F_{8.0 \mu m} > F_{5.8 \mu m}
 \end{aligned}&&&
   \label{eq:irac}
\end{flalign}
where $F_{3.6 \mu m}$, $F_{4.5 \mu m}$, $F_{5.8 \mu m}$ and $F_{8.0 \mu m}$ represent flux densities at 3.6, 4.5, 5.8 and 8.0 microns, respectively. We decided on this classification scheme because these criteria have been mostly developed using non-local, obscured galaxies, allowing us to avoid any `cosmic shift' in the IRAC colour-colour plot. With these criteria, 20 galaxies are classified as AGNs, while the remaining 326 are expected to be SF. The colour-colour plot and the location of all galaxies in the OB-I diagram can be seen in top and bottom panels of Fig. \ref{fig:appendix 3dhst}, respectively.

\subsection{MOSDEF}

In the MOSDEF data, we have all the necessary emission line fluxes to plot the NII diagram and compare it with the OB-I diagram. However, since we are working with redshifts above 1, we cannot use the classification schemes of \cite{kewley_2001} and \cite{kauffmann_2003}, as they were optimised for the lower redshifts. In order to take into account the `cosmic shift', we base ourselves on the theoretical work of \cite{kewley_2013} to separate galaxies between SF and AGN, as a function of redshift:

\begin{multline}
    \log\left(\frac{\oiii}{\hbeta}\right) = \frac{0.61}{\log(\nii/\halpha) - 0.02 - 0.1833z} \\ + 1.2 + 0.03z
    \label{eq:kew13sf}
\end{multline}

\begin{figure}[h]
    \centering
    \includegraphics[width=1.0\linewidth]{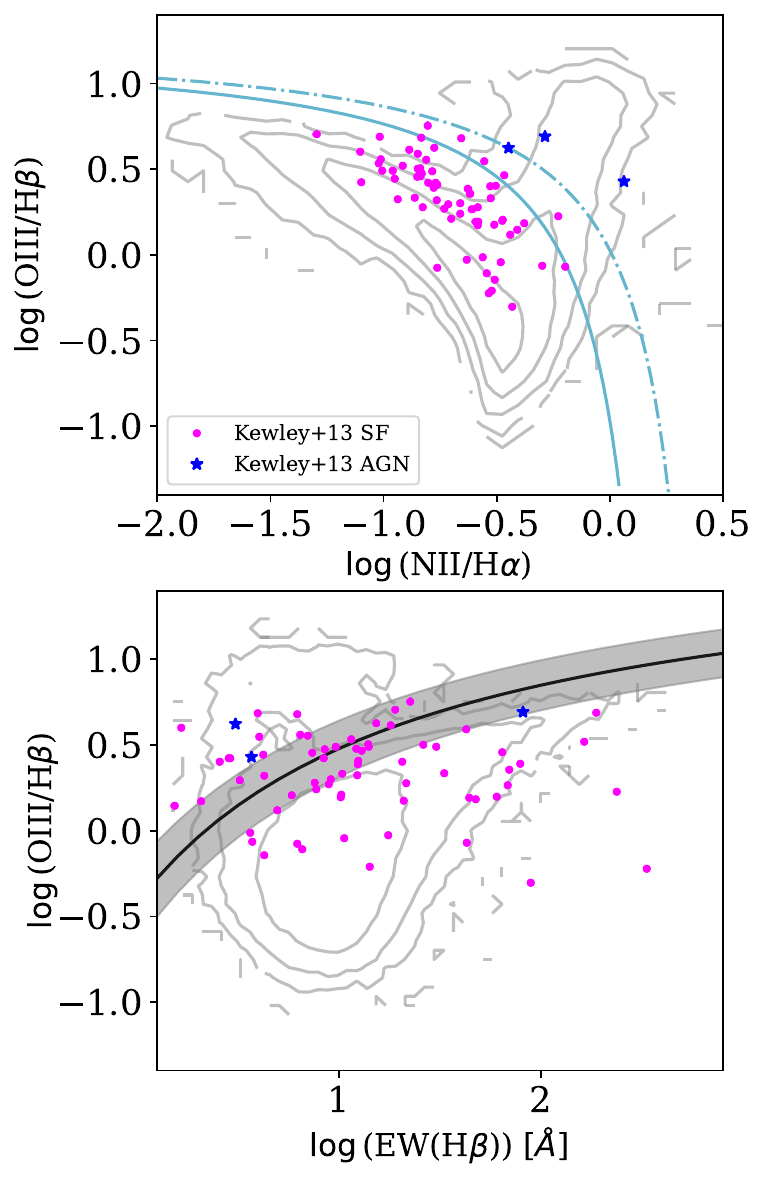}
    \caption{MOSDEF sample in the NII and OB-I diagrams. \textit{Top:} NII diagram. The full and dashed cyan lines represents the lowest and highest redshifts values from Eq. \ref{eq:kew13sf}, respectively, for the galaxies in the MOSDEF sample. The magenta dots represent SF galaxies and the blue stars AGNs according to the aforementioned classification. The grey contours represent the FADO-SDSS sample. \textit{Bottom:} OB-I diagram. The magenta dots and blue stars represent the same as in the NII diagram. The grey contours represents the FADO-SDSS sample and the black line and grey shaded area represent our empirical separation given by Eq. \ref{eq:hbeta limit}.}
    \label{fig:appendix mosdef}
\end{figure}

We took each galaxy present in the MOSDEF sample and applied this equation, to figure out if they are SF or AGN. With this classification scheme, we find that 3 galaxies are classified as AGN, while the remaining 67 are expected to be SF. The positions of these galaxies in the NII and OB-I diagrams can be seen in Fig. \ref{fig:appendix mosdef}.

\end{appendix}

\end{document}